\documentclass{article}
\usepackage{amsthm}
\usepackage{geometry}

    \newgeometry{vmargin={20mm}, hmargin={25mm,25mm}}
\newtheorem{theorem}{Theorem}{\bfseries}
{\bfseries}
\newtheorem{proposition}{Proposition}{\bfseries}
\newtheorem{corollary}{Corollary}{\bfseries}
\newtheorem{definition}{Definition}{\bfseries}
\newtheorem{remark}{Remark}{\itshape\bfseries}

\usepackage{graphicx}
\usepackage{subcaption}
\usepackage{amssymb}
\usepackage{float}
\usepackage{amsmath}
\usepackage{lscape}
\usepackage{braket}
\usepackage[colorlinks=true,linkcolor=blue,allcolors=blue]{hyperref}  
\usepackage{cite}
\newcommand{\pa}{\partial}
\newcommand{\opa}{\overline{\partial}}
\newcommand{\be}{\begin{equation}}
\newcommand{\ee}{\end{equation}}

\begin{document}

\title{\textbf{General solution of the exceptional Hermite differential equation and its minimal surface representation}}
\date{\vspace{-5ex}}
\maketitle

\begin{center}
V. Chalifour
\\
D{\'e}partement de Math{\'e}matiques et de Statistique, Universit\'{e} de Montr\'{e}al,\\
C. P. 6128, Succ. Centre-ville, Montr{\'e}al, Qu{\'e}bec, H3C 3J7, Canada\\
chalifour@dms.umontreal.ca
\\~\\
A.~M. Grundland
\\
1. Centre de Recherches Math{\'e}matiques, Universit{\'e} de Montr{\'e}al,\\
C. P. 6128, Succ. Centre-ville, Montr{\'e}al, Qu\'{e}bec, H3C 3J7, Canada\\ 2. D{\'e}partement de Math{\'e}matiques et Informatique, Universit\'{e} du Qu\'{e}bec,\\CP500, 
Trois-Rivi\`{e}res, Qu\'{e}bec, G9A 5H7, Canada\\
grundlan@crm.umontreal.ca
\\~\\
\vspace{.5in}
July 27 2020
\end{center}

\vspace{.5in}
\begin{abstract}
The main aim of this paper is the study of the general solution of the exceptional Hermite differential equation with fixed partition $\lambda = (1)$ and the construction of minimal surfaces associated with this solution. We derive a linear second-order ordinary differential equation associated with a specific family of exceptional polynomials of codimension two. We show that these polynomials can be expressed in terms of classical Hermite polynomials. Based on this fact, we demonstrate that there exists a link between the norm of an exceptional Hermite polynomial and the gap sequence arising from the partition used to construct this polynomial. We find the general analytic solution of the exceptional Hermite differential equation which has no gap in its spectrum. We show that the spectrum is complemented by non-polynomial solutions. We present an  implementation of the obtained results for the surfaces expressed in terms of the general solution making use of the classical Enneper-Weierstrass formula for the immersion in the Euclidean space $\mathbb{E}^3$, leading to minimal surfaces. Three-dimensional displays of these surfaces are presented.
\end{abstract}
\vspace{1in}

\newpage
\section{Introduction}
\label{sec:1}
Over the last decade, the problem of exceptional orthogonal polynomials (XOPs) has generated a great deal of interest and activity in several areas of mathematics and physics. Most of these activities focused on Jacobi and Laguerre XOPs (see \textit{e.g.} 
\cite{Bonneux2019,Bonneux2018a,Gomez-Ullate2010,Ho2012,Liaw2015,Odake2010a,Quesne2008,Quesne2012,Sasaki2010} 
and references therein), Hermite XOPs (see \textit{e.g.} \cite{Bonneux2018,Gomez-Ullate2014,Gomez-Ullate2018,Gomez-Ullate2013,Gomez-Ullate2016,Kuijlaars2015}) and multi-indexed orthogonal polynomials (see \textit{e.g.} 
\cite{Odake2013a,Odake2013b} and references therein). Exceptional orthogonal polynomials have been shown to play an essential role in several branches of physics, mostly related to the quantum harmonic oscillator. In particular, Jacobi XOPs were applied to the description of the Kepler-Coulomb quantum model \cite{Hoque2018} and were seen as having an electrostatic interpretation \cite{Dimitrov2014}. Hermite XOPs appeared first in the context of a time-independent Schr{\"o}dinger problem \cite{Dubov1992,Dubov1994,Samsonov1996}. They were also applied to the description of coherent states \cite{Hoffmann2018}.

The subject of study in our paper are exceptional Hermite polynomials. A substantial progress has been made recently in this area \cite{Bonneux2018,Felder2012,Kuijlaars2015,Gomez-Ullate2010,Gomez-Ullate2013,Gomez-Ullate2016,Gomez-Ullate2018,Milson2019}, among others by G{\'o}mez-Ullate et al \cite{Gomez-Ullate2014}, who provided a general formula for exceptional Hermite differential operators with an arbitrary partition. However, the associated ordinary differential equation (ODE) is a very complex one with a high degree of freedom and its general solution has yet to be established. In our study, we focus on a specific family of complex Hermite XOPs of codimension two, defined by the fixed partition $\lambda=(1)$. Our objective here is to formulate the differential equation corresponding to these polynomials, to find its general solution and to construct its geometric representation.

After determining the exceptional Hermite differential operator for a family of Hermite XOPs of codimension two with partition $\lambda=(1)$, we derive a linear second-order ODE associated with this family. We show that this ODE is equivalent to a non-degenerate confluent Heun equation. By the method of generalized series, we arrive at the general analytic solution of this ODE which has no gap in its spectrum. We demonstrate that the spectrum is supplemented by non-polynomial solutions in such a way that the Hermite XOPs of even and odd degrees correspond respectively to the even and odd part of a series which is part of the general solution. We show that these non-polynomial solutions arise naturally when considering the extension of the parameter of classical Hermite polynomials to negative integers.

Finally, we present a geometric representation and visualization of the obtained solution. The minimal surfaces are a natural choice here. The exceptional Hermite differential equation is a linear second-order ODE - a type which always admits a minimal surface representation, as shown in \cite{Doliwa2012}. We have also been motivated by the results of our previous work \cite{Chalifour2019} on the differential equations for classical orthogonal polynomials (including Hermite ones), in which explicit closed formulas for minimal surfaces were obtained. 

We construct minimal surfaces through the use of the soliton surface technique based on the Enneper-Weiertrass representation \cite{enneper1868analytisch,weierstrass1866fortsetzung}. For this purpose we express the Gauss-Weingarten equations as a linear system describing the moving frame attached to these surfaces. We identify this system with the exceptional Hermite ODE. We find the explicit form of the holomorphic wavefunction related to the linear spectral problem. Next we derive the explicit formula for the postulated minimal surfaces which is expressed in terms of the hypergeometric and error  functions. Given the very involved form of the obtained analytic expressions, the visualization of these surfaces has proven quite useful, revealing some properties of the solutions like symmetry features and singularities.

The paper is organized as follows. In section \ref{sec:2}, we recall some basic notions and definitions from the theory of Hermite XOPs. In section \ref{sec:3}, we derive the general solution of the exceptional Hermite differential equation with the fixed partition $\lambda=(1)$. In section \ref{sec:4}, we construct the minimal surfaces associated with the obtained general solution and illustrate them by numerical representations for different values of the parameter of the exceptional Hermite ODE.
\newpage
\section{Exceptional Hermite polynomials}
\label{sec:2}
\subsection{Sturm-Liouville operator in terms of Hermite polynomials}
To make the paper self-contained, we present in this section some known results concerning Hermite XOPs which are relevant for our purposes. A summary of recent developments on this subject can be found in the work of G\'omez-Ullate et al \cite{Gomez-Ullate2014}. As a starting point, consider the Sturm-Liouville problem
\be\label{eq:Sturm_Liouville}
L\psi = \lambda\psi
\ee
for the Schr{\"o}dinger operator possessing a potential $U(z)$
\be\label{eq:SchrodingerOp}
L = -\frac{d^2}{dz^2}+U(z).
\ee
If the operator (\ref{eq:SchrodingerOp}) is without monodromy, then the potential is of the form \cite{Oblomkov1999}
\be\label{eq:potential}
U(z) = -2\frac{d^2}{dz^2} \log{Wr(H_{k_1},H_{k_2},..., H_{k_l})}+z^2.
\ee
In this context, $\{k_i\}_{i = 1}^l$ is a strictly increasing sequence of positive integers and  $H_n(z)$ is the $n^{\text{th}}$ classical Hermite polynomial, which can be described by the Rodrigues formula \cite{Nikiforov1988}
\be\label{eq:Rodriques}
H_n(z) = (-1)^ne^{z^2}\frac{d^n}{dz^n}e^{-z^2}.
\ee
The potential (\ref{eq:potential}) is rational and has singularities corresponding to the zeros of the Wronskian
\be\nonumber
Wr(H_{k_1},H_{k_2},..., H_{k_l})(z),
\ee
which have been studied in \cite{Felder2012,Kuijlaars2015}. The following theorem (formulated in a slightly different way in \cite{Gomez-Ullate2014}) summarizes the results obtained by Krein \cite{Krein1957} and Adler \cite{Adler1994} concerning the zeros of the Wronskian of the eigenfunctions of the problem (\ref{eq:Sturm_Liouville}) in a more general context (general eigenfunctions). In our formulation, the sequence $\{k_1, ..., k_l\}$ is expressed as a new sequence $\{0,1,..., N_0',N_1, ..., N_1', ......, N_s, ...N_s'\}$, in order to clarify its structure. Here, according to the notation used in \cite{Adler1994}, the symbol prime $'$ denotes the biggest positive integer of the block $\{N_r, ..., N_r'\}$, while $N_r$ denotes the smallest positive integer of the same block.
\begin{theorem}\label{th:1}  (Krein-Adler) Let $\phi_j$ be the eigenfunctions of a pure-point Sturm-Liouville operator $L = -\frac{d^2}{dx^2} +U$ defined on the real line
\be
L[\phi_j] = \lambda_j\phi_j, \qquad j = 0, 1, 2, ...,\qquad x\in(-\infty,\infty),
\ee
with suitable boundary conditions. The Wronskian Wr$(\phi_{k_1}, ..., \phi_{k_l})$ has no zero on the real line if and only if the sequence of distinct positive integers $\{k_1, ..., k_l\}$, when arranged in an ascending order, has the following structure
\be\label{eq:CondThOne}
\{0, 1, ..., N_0'\}\cup\{N_1, ..., N_1'\}\cup...\cup\{N_s, ..., N_s'\},
\ee
where $N_r'+1 < N_{r+1}$ for all $r = 0, ..., s-1$. Here, the block $\{0, 1, ..., N_0'\}$ may be absent and the blocks $\{N_r, ..., N_r'\}$ consist of an even number of terms when $r\geq1$.
\end{theorem}
Condition (\ref{eq:CondThOne}) means that the sequence is allowed to (but does not necessarily) begin with a sequence of arbitrary length, composed of consecutive positive integers starting with zero, followed by an arbitrary number of blocks of even length. The meaning of the inequality $N_r'+1 < N_{r+1}$ is that there is a gap greater than 1 between the biggest positive integer $N_r'$ of a block and the smallest positive integer $N_{r+1}$ of the next block. These results are used to construct Hermite XOPs.
\newpage
\subsection{Construction of exceptional Hermite polynomials}
The definition of Hermite XOPs begins with the choice of a partition $\lambda = (\lambda_1, ..., \lambda_l)$ for a specific positive integer $m\in\mathbb{N}$, which consists of a non-decreasing sequence 
\be\label{eq:partition1}
0\leq\lambda_1\leq\lambda_2\leq...\leq \lambda_l.
\ee
For a combinatorial interpretation of the partition associated with XOPs, see \textit{e.g.} \cite{Gomez-Ullate2018,Bonneux2018a,Bonneux2018,Bonneux2019,Felder2012}. The sequence (\ref{eq:partition1}) is a  partition of a unique positive integer $m$ if
\be
m=\sum_{k=1}^l\lambda_k.
\ee
A sequence of type (\ref{eq:partition1}) determines a strictly increasing sequence called a gap sequence \cite{Gomez-Ullate2013}, of the form
\be\label{eq:gapSeq}
0\leq k_1<k_2<...< k_l,
\ee
where
\be\label{eq:gapSeqRel}
 k_i = \lambda_i +i - 1.
\ee
\begin{definition}
A partition of length $l$ is called a double partition if $l$ is even and $\lambda_{2i-1} = \lambda_{2i}$ for all $i$.
\end{definition}

From an arbitrary partition $\lambda$, we define a double partition of length $2l$ by duplicating each term of the partition (\ref{eq:partition1})
\be\label{eq:partition2}
\lambda^2 = (\lambda_1, \lambda_1, \lambda_2, \lambda_2, ..., \lambda_l, \lambda_l).
\ee
The application of relation (\ref{eq:gapSeqRel}) to any double partition of the form (\ref{eq:partition2}) leads to a strictly increasing sequence of length $2l$
\be
\{k_1, k_1+1, k_2, k_2 +1 ..., k_l, k_l+1\}
\ee
which respects the structure (\ref{eq:CondThOne}) of Theorem \ref{th:1}.
\begin{definition}
An Adler partition is either a double partition or a double partition preceded by a sequence of zeros of arbitrary length.
\end{definition}

For each partition $\lambda$, we consider the Wronskians
\begin{align}\label{eq:wronskiens}
&H_\lambda := Wr(H_{k_1}, ..., H_{k_l}),\\\label{eq:wronskiens2temp}
&H_{\lambda, n}:= Wr(H_{k_1}, ..., H_{k_l}, H_n), \quad n\notin\{k_1, ..., k_l\}.
\end{align}
\begin{definition}\label{def:polExc}
\cite{Gomez-Ullate2014} For any double partition (\ref{eq:partition2}) of length $2l$, we define the X$_\lambda$-Hermite family of polynomials, denoted by $\{H^{(\lambda)}_n\}$, as the following countable sequence
\be\label{def:HermiteX}
H^{(\lambda)}_n = H_{\lambda^2, n},\quad n\in\mathbb{N}\backslash\{k_1, k_1+1, k_2, k_2+1 ..., k_l, k_l+1\}.
\ee
\end{definition}
\begin{definition}\label{def:Codimension}\cite{Gomez-Ullate2010} 
Let 
$\lambda=(\lambda_1, \lambda_2, ..., \lambda_l)$
 be a partition of length $l$. The positive integer
\be
2m:=\left|\lambda^2\right|=\sum_{i=1}^l(\lambda_{2i-1}+\lambda_{2i}) = 2\sum_{i=1}^l \lambda_{2i}
\ee
 is called the codimension of the $X_\lambda$-Hermite family of polynomials. 
 \end{definition}
 
It is known \cite{Gomez-Ullate2014} that Hermite XOPs exist only for even codimension $2m$. This is a consequence of Theorem \ref{th:1} that leads to the choice of a double partition, as in equation (\ref{eq:partition2}).  In what follows, the notation $X_{2m}^{\lambda}$-Hermite will be used to refer to the family of Hermite XOPs associated with a partition $\lambda$ of some positive integer $m$ with codimension $2m$.

From the Wronskian (\ref{eq:wronskiens2temp}) and from Definition \ref{def:polExc}, we obtain \cite{Gomez-Ullate2014} that
\be\label{eq:degre}
deg H^{(\lambda)}_n(x) = 2\sum_{k=1}^l \lambda_k - 2l+n.
\ee
Equation (\ref{eq:degre}) tells us that the degree of a polynomial of the $X_{2m}^{\lambda}$-Hermite family is $n$ if and only if $\lambda_1>0$ and $m=l$, \textit{i.e.} the positive integer $m$ is equal to the length of its partition, which leaves only one possibility for the $m$-components partition, namely $\lambda = (1,1,...,1)$.
\subsection{Differential operator and orthogonality relation}
We consider the classical Hermite differential operator
\begin{align}\label{eq:HermioteOp}
T[y]:= \frac{d^2y}{dx^2} -2x \frac{dy}{dx}.
\end{align}
The exceptional Hermite operator is obtained through the use of state-deleting Darboux-Crum transformations and intertwining relations \cite{Gomez-Ullate2014} and making use of polynomial flags  \cite{Gomez-Ullate2013}
\begin{align}\label{eq:HermioteXOp}
T_\lambda[y]:=\frac{d^2y}{dx^2}-2\left(x+\frac{H_\lambda'}{H_\lambda}\right)\frac{dy}{dx}+\left(\frac{H_\lambda''}{H_\lambda}+2x\frac{H_\lambda'}{H_\lambda}\right)y,
\end{align}
where the symbol prime $'$ denotes the derivative with respect to $x$. Generally, the differential operator (\ref{eq:HermioteXOp}) has singular rational coefficients for an arbitrary partition $\lambda$. However, for an Adler partition $\lambda^2$, the operator $T_{\lambda^2}$ is non-singular on $\mathbb{R}$ and is called the X$_\lambda$-Hermite operator or exceptional Hermite operator \cite{Gomez-Ullate2014}. Oblomkov has studied regular singularities of the potential (\ref{eq:potential}) in \cite{Oblomkov1999}. We present below the results formulated in \cite{Gomez-Ullate2014}.
\begin{proposition}\cite{Gomez-Ullate2014} 
For every partition $\lambda$, we have
\be
T_\lambda[H_{\lambda, n}] = 2(l-n)H_{\lambda, n},\quad n\notin \{k_1, ..., k_l\}
\ee
where $l$ is the length of the partition.
\end{proposition}
\begin{corollary}\label{eq:EDOXHermite}\cite{Gomez-Ullate2014} 
The Hermite XOPs $H^{(\lambda)}_n$ introduced in Definition \ref{def:polExc} are eigenfunctions of the following second-order differential operator
\be\label{eq:OperatorTlambdaSquared}
T_{\lambda^2}[H^{(\lambda)}_n] = 2(2l-n)H^{(\lambda)}_n, \quad n\in\mathbb{N}\backslash \{k_1, k_1+1, ..., k_l, k_l+1\},
\ee
where $T_\lambda$ is given by (\ref{eq:HermioteXOp}).
\end{corollary}

If we define the polynomial of degree $l$
\be\label{eq:polyn}
p_\lambda(x):= (x-k_1)(x-k_2)\cdots(x-k_l),
\ee
then for any double partition $\lambda^2$, we have that 
\be
p_{\lambda^2}(n)>0\;\forall n\in\mathbb{N}\backslash \{k_1, k_1+1, ..., k_l, k_l+1\}.
\ee
\begin{proposition} \label{prop:orth} \cite{Gomez-Ullate2018} 
The Hermite XOPs $H^{(\lambda)}_n$ satisfy the orthogonality relation
\be\label{eq:Orthog_general}
\int_{-\infty}^{+\infty}H^{(\lambda)}_m(x)H^{(\lambda)}_n(x)W_{\lambda^2}(x)dx = \delta_{m,n}2^{n+2l}n!\sqrt{\pi}p_{\lambda^2}(n),
\ee
where the orthogonality weight given by
\be\label{eq:poids}
W_{\lambda^2}(x)=\frac{e^{-x^2}}{(H_{\lambda^2}(x))^2}>0
\ee
is regular.
\end{proposition}
\newpage
\section{General solution of the exceptional Hermite differential equation associated with the partition $\lambda = (1)$}
\label{sec:3}
In this section, we fix a partition $\lambda$ and use the theoretical results from section \ref{sec:2} to obtain the $X_{2m}^\lambda$-Hermite ODE associated with the chosen partition. We express the $X_{2m}^\lambda$-Hermite polynomials in terms of classical Hermite polynomials and we find the general solution of the $X_{2m}^\lambda$-Hermite ODE associated with the fixed partition.
\subsection{Exceptional Hermite polynomials in terms of classical and probabilistic Hermite polynomials}
The partition (\ref{eq:partition1}) can start with a sequence of zeros of arbitrary length.  In what follows, we consider a reduced double partition \cite{Gomez-Ullate2014}, \textit{i.e.} a partition for which $\lambda_1 >0$. If we set $m = 1$, then the only possible reduced partition is $\lambda = (1)$.  Therefore, $l=1$ and $\lambda^2 = (1,1)$ is a reduced double partition for which the associated strictly increasing sequence of length $2l$ is of the form $\{k_1, k_2\}$ (the gap sequence). From relation (\ref{eq:gapSeqRel}), we obtain
\begin{align}\label{eq:gapseqfixed}
k_1 = \lambda_1 +1 - 1 = 1,\qquad
k_2 = \lambda_2+2 - 1 = 2.
\end{align}
Due to Definition \ref{def:Codimension}, the codimension of the family of polynomials which results from this choice of partition is $2m = 2$.  We therefore consider the countable family of polynomials which consitutes the $X_2^{(1)}$-Hermite family $\{H_n^{(1)}(x)\;|\; n\in\mathbb{N}\backslash \{1, 2\}\}$. From relation (\ref{eq:degre}), we see that the degree of a polynomial of this family reduces to
\be\label{eq:Degre2014}
deg H^{(1)}_n(x) = n\quad \forall n\in\mathbb{N}\backslash \{1, 2\}.
\ee
The Wronskians defined in (\ref{eq:wronskiens}) and (\ref{eq:wronskiens2temp}) become
\begin{align}\label{eq:polyn_poids_1}
&H_{(1,1)}(x) = Wr(H_{1}, H_{2})(x) = 4(1+2x^2),\\\label{eq:HermiteExceptFixe}
&H^{(1)}_n(x) = H_{(1,1), n}(x)= Wr(H_{1}, H_{2}, H_n),
\end{align}
where $n\notin\{1, 2\}$. Under the above assumptions, we obtain the following result.\\
\begin{theorem}\label{cor:Poynome2019}
For the fixed partition $\lambda = (1)$, the polynomials (\ref{eq:HermiteExceptFixe}) satisfy the relation
\be\label{eq:corol_rec}
H_{n}^{(1)}(x) = 8(n-1)(n-2)\hat{H}_{n}(x),\qquad \forall n\in\mathbb{N}\backslash\{1, 2\}
\ee
where $\hat{H}_{n}(x)$ is defined as
\be\label{eq:HChapeau}
\hat{H}_{n}(x):= H_n(x)+4nH_{n-2}(x)+4n(n-3)H_{n-4}(x).
\ee
\end{theorem}
\begin{proof}
Making use of the differential relation \cite{Szego1939}
\be\label{eq:HermiteDerivee}
H_n'(x) = 2nH_{n-1}(x)
\ee
and Definition \ref{def:polExc}, we find
\begin{align}
H_{n}^{(1)}
&=16\left(H_n -2nxH_{n-1}+2n(n-1)x^2H_{n-2}+n(n-1)H_{n-2}\right).\label{eq:temp1}
\end{align}
Through successive applications of the following recurrence relation \cite{Slavyanov2000}
\be\label{eq:recurrHermiteClassique}
2xH_{n+1}(x) = 2(n+1)H_n(x)+H_{n+2}(x)
\ee
to equation (\ref{eq:temp1}), we obtain
\be
H_{n}^{(1)}= 8(n-1)(n-2)\left(H_n(x)+4nH_{n-2}(x)+4n(n-3)H_{n-4}(x)\right).
\ee
\end{proof}
$\left.\right.\hfill\square$
\begin{remark}
Expression (\ref{eq:HChapeau}) for the function $\hat{H}_n(x)$ appeared first in a slightly different form under the notation $F_n(x)$ in \cite{Dubov1992}. In this remark, we establish the correspondence between this function and the function $\hat{H}_n(x)$ (\ref{eq:HChapeau}), leading to an alternative description of $X_2^{(1)}$-Hermite polynomials in terms of probabilistic Hermite polynomials. 
The function $F_n(x)$ arises in the context of a time-independent Schr{\"o}dinger problem
\be\label{eq:ScrodingerDubov1992}
H\psi_E = E\psi_E
\ee
involving the linear differential operator
\be
H = -\frac{1}{2}\frac{d^2}{dx^2}+\frac{1}{8}x^2+\frac{2}{1+x^2}-\frac{4}{(1+x^2)^2}+\frac{1}{3}.
\ee
Under the assumption that the eigenfunctions are of the form
\be
\psi_{E} = \frac{F(x)}{1+x^2}e^{-x^2/4},
\ee
equation (\ref{eq:ScrodingerDubov1992}) is equivalent to the differential equation for the unknown function $F_n(x)$
\be\label{eq:ScrodingerDubov1992Equiv}
(1+x^2)\left( F_n''(x)-xF_n'(x)+\epsilon_n F_n(x)\right) = 4xF_n'(x), \qquad \epsilon_n = 2(E_n+5/12).
\ee
The polynomial solutions of equation (\ref{eq:ScrodingerDubov1992Equiv}) are given \cite{Dubov1992} by
\begin{align}\label{eq:PolDubov1992}
F_0(x) &= 1,\qquad\qquad\qquad\qquad\qquad\qquad\qquad\qquad\qquad\qquad\qquad\qquad\;\;\;\; \epsilon_0 = 0,\qquad\;\;\; n=0,\\\nonumber
 F_n(x) &= He_{n+2}(x)+2(n+2)He_{n}(x) +(n+2)(n-1)He_{n-2}(x), \qquad \epsilon_n = n+2,\quad n\geq1,
\end{align}
where $He_{n}(x)$ is the $n^{\text{th}}$ probabilistic Hermite polynomial defined \cite{Nikiforov1988} by
\be
He_{n}(x) = (-1)^n e^{x^2/2}\frac{d^n}{dx^n}e^{-x^2/2}.
\ee
We note that there is a gap in the spectrum of the eigenvalue problem (\ref{eq:ScrodingerDubov1992Equiv}) which corresponds to the gap sequence (\ref{eq:gapseqfixed}) associated with $X_2^{(1)}$-Hermite polynomials. Making use of the following relation \cite{Nikiforov1988}
\be
He_{n}(x)=2^{-n/2}H_n\left(\frac{x}{\sqrt{2}}\right),
\ee
we obtain the connection between $F_n(x)$ and $\hat{H}_n(x)$ and applying it to relation (\ref{eq:HChapeau}), we get
\begin{align}
&\hat{H}_0(x) = F_0\left(\sqrt{2}x\right)=1,\qquad\qquad\quad\; n=0,\\\nonumber
 &\hat{H}_{n+2}(x) = 2^{(n+2)/2}F_n\left(\sqrt{2}x\right),\qquad\;\, n\geq1.
\end{align}
The properties of the function $\hat{H}_n(x)$ (\ref{eq:HChapeau}) have also been discussed in \cite{Cariena2008}, where we find the analogue of the Rodrigues formula (\ref{eq:Rodriques}) for this case
\be
\hat{H}_n(x) = (-1)^ne^{x^2}\left(\frac{d^n}{dx^n}+4n \frac{d^{n-2}}{dx^{n-2}}+4n(n-3)\frac{d^{n-4}}{dx^{n-4}}\right)e^{-x^2}.
\ee
\end{remark}
\newpage
\subsection{Orthogonality relation of the $X_2^{(1)}$-Hermite polynomials}
When $\lambda = (1)$, the polynomial (\ref{eq:polyn}) associated with the double partition $\lambda^2=(1,1)$ becomes
\be\label{eq:polyn_1}
p_{(1,1)}(n) = (n-1)(n-2)\geq0 \qquad \forall \; n\in\mathbb{N}.
\ee
Using (\ref{eq:polyn_poids_1}) in relation (\ref{eq:poids}), we obtain the weight function
\be\label{eq:Poidslambda1}
W_{(1,1)}(x)=\frac{e^{-x^2}}{(H_{(1,1)}(x))^2} = \frac{e^{-x^2}}{(4(1+2x^2))^2}>0\quad\forall x\in\mathbb{R}.
\ee
The orthogonality relation (\ref{eq:Orthog_general}) becomes
\be\label{eq:orthogX1}
\int_{-\infty}^{+\infty}H^{(1)}_m(x)H^{(1)}_n(x)\frac{e^{-x^2}}{(4(1+2x^2))^2}dx = \delta_{m,n}\sqrt{\pi}2^{n+2}n!(n-1)(n-2),
\ee
where $m, n \in \mathbb{N}\backslash\{1,2\}$. Equivalently, using Theorem \ref{cor:Poynome2019}, we find
\be\label{eq:orthogPolynome2019}
\int_{-\infty}^{+\infty}\hat{H}_{m}(x)\hat{H}_{n}(x)\frac{e^{-x^2}}{(1+2x^2)^2}dx = \delta_{m,n}\frac{\sqrt{\pi}2^{n}n!}{(n-1)(n-2)},
\ee
which was shown independently in \cite{Cariena2008}. Because of an order relation between integrands, the integral (\ref{eq:orthogX1}) diverges if the integral (\ref{eq:orthogPolynome2019}) diverges. Indeed, for all $m = n\in\mathbb{N}\backslash\{1,2\}$, we have that $0<4^2\hat{H}_{n}^2<4^3(n-1)^2(n-2)^2\hat{H}_{n}^2$. 
Using Theorem \ref{cor:Poynome2019}, we find that $0<\hat{H}_{n}^2<(H^{(1)}_{n})^2/4^2$. Multiplying by $e^{-x^2}/(1+2x^2)^2$ and integrating on the orthogonality interval $(-\infty, +\infty)$, we obtain
\be
0<\int_{-\infty}^{+\infty}\hat{H}_{n}^2\frac{e^{-x^2}}{(1+2x^2)^2}dx<\int_{-\infty}^{+\infty}\left(H^{(1)}_n\right)^2\frac{e^{-x^2}}{(4(1+2x^2))^2}dx.
\ee
The norm of $X_2^{(1)}$-Hermite polynomials is therefore defined on $\mathbb{N}$ except for integer values which are zeros of the polynomial (\ref{eq:polyn_1}), namely $n = 1, 2$.  For $\lambda=(1)$, these integer values correspond to the gap sequence (\ref{eq:gapSeq}).
\subsection{$X_2^{(1)}$-Hermite differential equation and its link to a specific Heun equation}
Consider the first and second-order derivatives of the polynomial (\ref{eq:polyn_poids_1}) obtained from the double partition $\lambda^2 = (1,1)$
\be\label{eq:temp3a}
H_{(1,1)}(x)  = 4(1+2x^2),\qquad
H_{(1,1)}'(x) =16x,\qquad
H_{(1,1)}''(x) = 16.
\ee
Corollary \ref{eq:EDOXHermite}, written in terms of the differential operator (\ref{eq:HermioteXOp}), gives us
\begin{align}\label{eq:temp5}
&H_{\lambda^2, n}''-2\left(x+\frac{H_{\lambda^2}'}{H_{\lambda^2}}\right)H_{\lambda^2, n}'+\left(\frac{H_{\lambda^2}''}{H_{\lambda^2}}+2x\frac{H_{\lambda^2}'}{H_{\lambda^2}}\right)H_{\lambda^2, n} = (2l-4n)H_{\lambda^2, n}.
\end{align}
Making use of equations (\ref{eq:temp3a}) in (\ref{eq:temp5}) and simplifying, we obtain
\begin{align}
\left(H^{(1)}_n(x)\right)''-2\left(x+\frac{4x}{(1+2x^2)}\right)\left(H^{(1)}_n(x)\right)'+2nH^{(1)}_n(x) = 0.
\end{align}
In other words, the polynomial $H^{(1)}_n(x)$ (\ref{eq:HermiteExceptFixe}) is a solution of the second-order linear homogeneous ODE
\be\label{eq:EDOX}
\omega''(x)-2\left(x+\frac{4x}{1+2x^2}\right)\omega'(x)+2n\omega(x) = 0, \qquad n\in \mathbb{N}\backslash\{1,2\},\;\;x\in\mathbb{R},
\ee
which was presented in \cite{Milson2019} as the exceptional Hermite differential equation. In our further analysis, we consider the complex extension of the ODE (\ref{eq:EDOX})
\be\label{eq:EDOXComplexe}
\omega''(z)-2\left(z+\frac{4z}{1+2z^2}\right)\omega'(z)+2n\omega(z) = 0, \qquad n\in \mathbb{N},\;\; z\in\mathbb{C}.
\ee  
From this point on, we will refer to equation (\ref{eq:EDOXComplexe}) as the complex $X_2^{(1)}$-Hermite ODE. The linear second-order differential operator describing the left-hand side of equation (\ref{eq:EDOXComplexe}) is given by
\be\label{eq:operatorA}
A:=\frac{d^2}{dz^2} +p(z)\frac{d}{dz}+q(z),
\ee
where
\be\label{eq:coeff}
p(z) = -2\left(z+\frac{4z}{1+2z^2}\right),\qquad q(z) = 2n.
\ee
We consider the Laurent expansion around a finite point $z_0\in\mathbb{C}$ of $p(z)$ and $q(z)$
\begin{align}\label{eq:Laurentpfinite}
p(z) &= \sum_{k=-\infty}^{\infty}p_{z_0,k}(z-z_0)^k,\qquad |z-z_0|<r_p \text{ for some } r_p>0,\\\label{eq:Laurentqfinite}
q(z) &= \sum_{k=-\infty}^{\infty}q_{z_0,k}(z-z_0)^k,\qquad |z-z_0|<r_q \text{ for some } r_q>0,
\end{align}
as well as the asymptotic expansion associated with infinity $z_0 = \infty$
\begin{align}\label{eq:LaurentpInfinite}
p(z) &= \sum_{k=-\infty}^{\infty}p_{\infty,k}(z-z_0)^k,\qquad |z-z_0|>R_p \text{ for some } R_p\geq0,\\\label{eq:LaurentqInfinite}
q(z) &= \sum_{k=-\infty}^{\infty}q_{\infty,k}(z-z_0)^k,\qquad |z-z_0|>R_q \text{ for some } R_q\geq0.
\end{align}
We determine if the singularities of the operator $A$ (\ref{eq:operatorA}) are regular (Fuchsian) or irregular according to the definitions \cite{Derezinski2020} below.
\begin{definition}\label{def:degreeSingularity}
The degree of the singularity of $p$ at $z_0$ is defined by
\begin{align}
deg(p,z_0):&= - \mathrm{min}\{k\;|\;p_{z_0,k}\neq0\},\quad z_0\in\mathbb{C},\\
deg(p,\infty):&=  \mathrm{max}\{k\;|\;p_{\infty,k}\neq0\},\quad\;\; z_0=\infty,
\end{align}
where $p_{z_0,k}$ and $p_{\infty,k}$ are the $k^{\text{th}}$ coefficients of the Laurent expansion (\ref{eq:Laurentpfinite}) and of the asymptotic expansion (\ref{eq:LaurentpInfinite}), respectively.
\end{definition}
\begin{definition}\label{def:Regularpoint}
A point $z_0$ is a regular point of the operator $A$ (\ref{eq:operatorA}) if 
\begin{align}
&deg(p,z_0)\leq0,\qquad\qquad\quad\;\;\; deg(q,z_0)\leq0,\qquad\;\;\, z_0\in\mathbb{C},\\
&deg\left(p-\frac{2}{z},\infty\right)\leq-2,\qquad deg(q,\infty)\leq-4,\qquad z_0=\infty.
\end{align}
Otherwise, we say that $z_0$ is a singular point of $A$.
\end{definition}
\begin{definition}\label{def:Singularpoint}
A singular point $z_0$ is regular (Fuchsian) if 
\begin{align}
&deg(p,z_0)\leq1,\qquad \,\;deg(q,z_0)\leq2,\qquad\;\,\; z_0\in\mathbb{C},\\
&deg\left(p,\infty\right)\leq-1,\quad\; deg(q,\infty)\leq-2,\qquad z_0=\infty.
\end{align}
\end{definition}
From equations (\ref{eq:Laurentpfinite})-(\ref{eq:LaurentqInfinite}) and from Definition \ref{def:degreeSingularity}, we find
\begin{align}
&deg(p,z_0)= 1,\qquad\qquad\;\, deg(q,z_0) = 0 ,\quad\;\,\, z_0\in\left\{\pm i/\sqrt{2}\right\},\\
&deg\left(p-\frac{2}{z},\infty\right)=1 \qquad deg(q,\infty)=0,\quad\;\; z_0=\infty.
\end{align}
Using Definitions \ref{def:Regularpoint} and \ref{def:Singularpoint}, we conclude that the operator A (\ref{eq:operatorA}) possesses two regular singular points at $\{\pm i/\sqrt{2}\}$ and an irregular singular (non-Fuchsian) point at $\{\infty\}$.

Performing the transformation of the independent variable $z = \sqrt{t/2}$ in the complex $X_2^{(1)}$-Hermite ODE (\ref{eq:EDOXComplexe}) and multiplying it by $(1+t)/4$, we get
\be\label{eq:ConfHeunNormalForm}
2t(1+t)\omega''(t)-\left(t^2+4t-1\right)\omega'(t)+\frac{n}{2}(1+t)\omega(t) = 0.
\ee
Then, making the transformation $t = -s$ and dividing by $2s(s-1)$, we obtain a non-degenerate confluent Heun equation of the form
\be\label{eq:Heun}
\omega''(s)+\left(\frac{\gamma}{s}+\frac{\delta}{s-1}+\epsilon \right)\omega'(s)+\frac{\alpha s-q}{s(s-1)}\omega(s) = 0,
\ee
where
\be\nonumber
\gamma =\epsilon =  -\frac{1}{2},\quad \delta = 2,\quad \alpha =q= -\frac{n}{4}.
\ee
According to \cite{Filipuk2020,Ohyama2006}, there exists a correspondence between a specific Painlev{\'e} V equation and the confluent Heun equation (\ref{eq:Heun}) and thus, consequently, it exists also between the same Painlev{\'e} V equation and the complex $X_2^{(1)}$-Hermite ODE (\ref{eq:EDOXComplexe}).
\subsection{Main result - Polynomial and non-polynomial solutions of the $X_2^{(1)}$-Hermite differential equation}
In this section, we study the polynomial and non-polynomial solutions of the complex $X_{2}^{(1)}$-Hermite ODE (\ref{eq:EDOXComplexe}). We find new solutions using the method of generalized series. We show the proportionality relations between these solutions and the Hermite XOPs and we perform an extension of the classical Hermite polynomials to negative integers $n$, leading to non-polynomial solutions.
\begin{corollary}\label{cor:HChapeau}
The function $\hat{H}_n(x)$ defined in (\ref{eq:HChapeau}) is a polynomial solution of the ODE (\ref{eq:EDOX}) for all $n\in\mathbb{N}\backslash\{1,2\}$. Moreover, this solution and the polynomial $H^{(1)}_n(x)$ defined in (\ref{eq:HermiteExceptFixe}) are proportional to each other for all $n\in\mathbb{N}\backslash\{1,2\}$.
\end{corollary}
\begin{proof}
Let $n\in\mathbb{N}\backslash\{1,2\}$. The proof is straightforward, considering that the polynomial $\hat{H}_n(x)$ is equal to the polynomial $H^{(1)}_n(x)$, up to a constant (which depends on $n$), by Theorem \ref{cor:Poynome2019}.
\end{proof}
\begin{remark}
By Corollary \ref{cor:HChapeau}, we know that the polynomials $H^{(1)}_n(z)$ and $\hat{H}_n(z)$ are linearly dependent solutions of the ODE (\ref{eq:EDOXComplexe}) for $n\in\mathbb{N}\backslash\{1,2\}$.  However, given the relation (\ref{eq:HermiteExceptFixe}) and Theorem \ref{cor:Poynome2019}, we see that $H^{(1)}_n(z)$ is a trivial solution of the ODE (\ref{eq:EDOXComplexe}) for $n=1,2$.
\\~\\
In what follows, we construct a non-trivial and non-polynomial solution of the ODE (\ref{eq:EDOXComplexe}) for $n=1, 2$. We make use of the extension of the parameter $n\in\mathbb{N}$ of the classical Hermite polynomial $H_n(z)$ to negative integers $n=-1,-2,-3$. Using the Rodrigues formula (\ref{eq:Rodriques}), we define
\be
H_{-1}(z) := \frac{\sqrt{\pi}}{2}e^{z^2} \left(1-\mathrm{erf}(z)\right),\quad n=-1,
\ee
where $\mathrm{erf}(z)$ is the error function \cite{Nikiforov1988} defined by
\be\label{eq:temp17}
\mathrm{erf}(z) = \frac{2}{\sqrt{\pi}}\int_0^z e^{-t^2}dt.
\ee
Making use of the recurrence relation (\ref{eq:recurrHermiteClassique}), we find
\begin{align}\label{eq:CalculsHNeg}
H_{-1}(z) &= \frac{\sqrt{\pi}}{2}e^{z^2} \left(1-\mathrm{erf}(z)\right),\qquad \qquad\qquad\qquad\quad\,\,\; n=-1,\\\nonumber
H_{-2}(z) &=\frac{1}{2}\left(1-e^{z^2}\sqrt{\pi}z(1-\mathrm{erf}(z))\right),\quad\qquad\qquad\;\;\, n=-2,\\\nonumber
H_{-3}(z) &=\frac{1}{8}\left(-2z+e^{z^2}\sqrt{\pi}(1+2z^2)(1-\mathrm{erf}(z))\right),\quad n=-3.
\end{align}
Substituting equations (\ref{eq:CalculsHNeg}) into (\ref{eq:HChapeau}), we obtain
\begin{align}\label{eq:HermiteNegInt}
\hat{H}_1(z) &= 4z+\sqrt{\pi} e^{z^2}(1-2z^2)\left(1-\mathrm{erf}(z)\right),\qquad
\hat{H}_2(z) =2+4z^2+4\sqrt{\pi} ze^{z^2}\left(1-\mathrm{erf}(z)\right).
\end{align}
The functions (\ref{eq:HermiteNegInt}), as can be easily shown, constitute non-polynomial and nontrivial solutions of the complex $X_{2}^{(1)}$-Hermite ODE (\ref{eq:EDOXComplexe}) for $n=1,2$.
\end{remark}
\begin{remark}
Let us note that the extension of classical Hermite polynomials to negative integers $n$ can be performed numerically by the \textit{WolframAlpha} application.
\end{remark}

We now use the method of generalized series to construct the general solution of the complex $X_{2}^{(1)}$-Hermite ODE (\ref{eq:EDOXComplexe}). The variable coefficients $p(z)$ and $q(z)$ given by (\ref{eq:coeff}) are analytic on $B_{\delta}(0)$, where $\delta = 1/\sqrt{2}$. Therefore, we can find at least one solution by the method of generalized series. Moreover, the principal part of the Laurent series around $z_0 = 0$ must vanish. Let us introduce the series
\be\label{eq:Series1and2}
\beta_n(z) = z^{\sigma_1}\sum_{k=0}^\infty c_k(n) z^k ,\qquad \nu_n(z) = z^{\sigma_2}\sum_{k=0}^\infty \tilde{c}_k(n) z^k,\quad n\in\mathbb{N},
\ee
where $\sigma_1$ and $\sigma_2$ are the roots of the indicial equation associated with equation (\ref{eq:EDOXComplexe})
\be\label{eq:EQDet}
\sigma(\sigma - 1)+a_0\sigma +b_0 = 0.
\ee
By definition, we have
\be
a_0 = \lim_{z\rightarrow 0}z\cdot\left(-2z-\frac{8z}{1+2z^2}\right) = 0,\quad b_0 = \lim_{z\rightarrow 0}z^2\cdot2n = 0.
\ee
The indicial equation (\ref{eq:EQDet}) reduces to 
\be
\sigma(\sigma - 1) = 0,
\ee
which possesses the roots $\sigma_1 = 0$ and $\sigma_2 = 1$. The two series (\ref{eq:Series1and2}) become
\begin{align}\label{eq:SerieRacine1}
\beta_n(z)  &= \sum_{k=0}^\infty c_k(n) z^k,\\\label{eq:SerieRacine2}
\nu_n(z)  &= \sum_{k=0}^\infty \tilde{c}_k(n) z^{k+1}.
\end{align}
Case 1: root $\sigma_1 = 0$. 
We substitute the series $\beta_n(z)$ (\ref{eq:SerieRacine1}) and its derivatives up to order two into the ODE (\ref{eq:EDOXComplexe}) and obtain
\begin{align}\nonumber
(2c_2 +2nc_0) + &(6c_3+2(n-5)c_1)z \\
&+\sum_{k=2}^\infty\left[(k+2)(k+1)c_{k+2}+2(k(k-6)+n)c_k+4(n-k+2)c_{k-2}\right]z^k = 0,
\end{align}
where $c_0$ and $c_1$ are arbitrary constants. Let $c_0 = c_1 = 1$. Then the first coefficients take the form
\be
c_2  = -n, \qquad\quad c_3 = -\frac{1}{3}(n-5), \qquad\quad c_4 = \frac{1}{6}n(n-10),
\ee
and we conclude that for all $k\geq4$, the recurrence relation is as follows
\be\label{eq:recSerie}
c_k = \frac{-2((k-2)(k-8)+n)c_{k-2}-4(n-k+4)c_{k-4}}{k(k-1)}.
\ee
The first even and odd coefficients are presented in Table \ref{tab:1}.
\begin{table}[H]
\centering
\caption{First coefficients of the series $\beta_n(z)$}
\begin{tabular}{lll}
\hline\noalign{\smallskip}
$\quad k$ & $\quad c_{2k}(n)$ & $\quad c_{2k-1}(n)$  \\
\noalign{\smallskip}\hline\noalign{\smallskip}
$\quad0$     &$\quad1$  &  \quad-   \\
$\quad1$     & $\quad-n$  &  $\quad1$   \\
$\quad2$     &  $\quad\frac{n^2-10n}{6}$  & $ \quad-\frac{n-5}{3}$    \\
$\quad3$&  $\quad-\frac{ n^3-30 n^2+104 n }{90}$  &  $\quad\frac{n^2-20n+51}{30}$     \\
$\quad4$&  $\quad\frac{       + n^4- 60 n^3+ 524 n^2-1200 n}{2520}$ & $\quad-\frac{  n^3- 45 n^2+311 n-555 }{630}$     \\
$\quad5$&$\quad- \frac{              n^5-100 n^4+1580 n^3-8720 n^2+ 15744 n}{113400}$   &  $\quad\frac{        n^4- 80 n^3+ 1046 n^2- 4720 n+6825  }{22680}    $\\
\noalign{\smallskip}\hline
\end{tabular}
\label{tab:1}       
\end{table}
\noindent For $k\geq2$, the denominators of the even coefficients $c_{2k}$ from Table \ref{tab:1} take the form $(2k)!/2^{k}$ while the denominators of the odd coefficients $c_{2k-1}$ from Table \ref{tab:1} take the form $(2k-1)!/2^{k-1}$. 
The sign of the highest power of $n$ appearing in the even and odd coefficients from Table \ref{tab:1} alternates. Therefore, the coefficients of the series $\beta_n(z)$ (\ref{eq:SerieRacine1}) are of the form
\be\label{eq:CoeffTemp}
c_{2k}(n) = (-1)^k\frac{p_k(n)}{(2k)!/2^{k}},\qquad c_{2k-1}(n) = (-1)^{k+1}\frac{q_k(n)}{(2k-1)!/2^{k-1}},
\ee
where $p_k(n)$ and $q_k(n)$ are polynomials of the positive integer variable $n$ of order $k$ and $k-1$, respectively. We denote by $\lambda_p(k)$ and $\lambda_q(k)$ the roots of the polynomials $p_k(n)$ and $q_k(n)$, respectively.  
\begin{table}[H]
\caption{Roots of the coefficients of the series $\beta_n(z)$}
\label{tab:2}       
\centering
\begin{tabular}{lll}
\hline\noalign{\smallskip}
$\quad k$ & $\qquad \lambda_p(k)$ & $\qquad \lambda_q(k)$  \\
\noalign{\smallskip}\hline\noalign{\smallskip}
$\quad0$     & \qquad-  &  \qquad-   \\
$\quad1$     & $\qquad0$  &  \qquad-   \\
$\quad2$     &  $\qquad0,\mathbf{10}$  & $\qquad\mathbf{5}$   \\
$\quad3$& $\qquad0,4,\mathbf{26}$   & $\qquad3,\mathbf{17}$     \\
$\quad4$&  $\qquad0,4,6,\mathbf{50}$  & $\qquad3,5,\mathbf{37}$     \\
$\quad5$&    $\qquad0,4,6,8,\mathbf{82}$  &  $\qquad3,5,7,\mathbf{65}\quad$     \\
\noalign{\smallskip}\hline
\end{tabular}
\end{table}
\noindent Table \ref{tab:2} shows the roots of the first terms associated with the series $\beta_n(z)$ (\ref{eq:SerieRacine1}). For $k\geq3$, the coefficients $c_{k}$ (even and odd) have in particular as a root $\lambda(k) = (k-1)^2+1$. These roots correspond to the positive integers in bold character.  The coefficients $c_{2k}$ then possess the factor
\be\label{eq:Factor1}
(n-((2k-1)^2+1)),
\ee
while the coefficients $c_{2k-1}$ possess the factor
\be\label{eq:Factor2}
(n-((2(k-1))^2+1)).
\ee
If the coefficient is even, the remaining roots consist of a sequence of even positive integers
$0, 4, 6, ..., k-2$, where the positive integer $2$ is excluded. The even coefficients then possess the factors
\be\label{eq:Factor3}
n\cdot\prod_{j=1}^{k-2}(n-2(1+j)).
\ee
If the coefficient is odd, the remaining roots consist of a sequence of odd positive integers $3, 5, 7 ..., k-2 $, where the positive integer $1$ is excluded. The odd coefficients then possess the factors
\be\label{eq:Factor4}
\prod_{j=1}^{k-2}(n-2(1+j)+1).
\ee
Fixing the positive integer $n$ therefore truncates the series of even or odd coefficients, but not both, depending on the parity of $n$.  Since even and odd coefficients have no root in common, the series must be infinite. Moreover, the fact that the positive integers $1$ and $2$ are excluded indicates that the $X_2^{(1)}$-Hermite family of polynomials is defined on the spectrum  $\mathbb{N}\backslash\{1,2\}$.

Making use of (\ref{eq:Factor1})-(\ref{eq:Factor4}), we find
\begin{align}\nonumber
p_2(n) &=n(n-10),\qquad\qquad\qquad\qquad\qquad\qquad\qquad\qquad\, k=2,\\\label{eq:pk}
p_k(n) &= n(n-((2k-1)^2+1))\prod_{j=1}^{k-2}(n-2(1+j)),\quad\quad\;\;\; k\geq3,\\\nonumber
q_2(n)& = (n-5),\qquad\qquad\qquad\qquad\qquad\qquad\qquad\qquad\;\;\;\;\, k=2,\\\label{eq:qk}
 q_k(n) &= (n-((2(k-1))^2+1))\prod_{j=1}^{k-2}(n-2(1+j)+1),\quad k\geq3,
\end{align}
and taking (\ref{eq:pk}) and (\ref{eq:qk}) into account, we obtain the coefficients (\ref{eq:CoeffTemp}) in the form
\begin{align}\label{eq:coeffPair}
c_{2k}(n) &= (-1)^k \frac{n(n-((2k-1)^2+1))\prod_{j=1}^{k-2}(n-2(1+j))}{(2k)!/2^{k}},\\\label{eq:coeffImpair}
 c_{2k-1}(n) &=(-1)^{k+1} \frac{(n-((2(k-1))^2+1))\prod_{j=1}^{k-2}(n-2(1+j)+1)}{(2k-1)!/2^{k-1}}.
\end{align}
The series $\beta_n(z)$ (\ref{eq:SerieRacine1}) therefore takes the form
\begin{align}\nonumber
&\beta_n(z) = 1+z-nz^2-\frac{n-5}{3}z^3+\frac{n(n-10)}{6}z^4\\\label{eq:Sol1}
&\qquad\quad+\sum_{k=3}^\infty \left[(-1)^k \frac{n(n-((2k-1)^2+1))\prod_{j=1}^{k-2}(n-2(1+j))}{(2k)!/2^{k}}z^{2k}\right.\\\nonumber
&\left.\qquad\quad +(-1)^{k+1} \frac{(n-((2(k-1))^2+1))\prod_{j=1}^{k-2}(n-2(1+j)+1)}{(2k-1)!/2^{k-1}}z^{2k-1}\right].
\end{align}
The series $\beta_n(z)$ (\ref{eq:Sol1}) converges. Indeed, if the series converges absolutely, then it may be written in terms of two separate series for the even and for the odd coefficients
\begin{align}\nonumber
&\beta_n(z) = 1+z-nz^2-\frac{n-5}{3}z^3+\frac{n(n-10)}{6}z^4\\\label{eq:s1}
&\qquad\quad+\sum_{k=3}^\infty \left[(-1)^k \frac{n(n-((2k-1)^2+1))\prod_{j=1}^{k-2}(n-2(1+j))}{(2k)!/2^{k}}z^{2k}\right]\\\label{eq:s2}
& \qquad\quad+\sum_{k=3}^\infty\left[(-1)^{k+1} \frac{(n-((2(k-1))^2+1))\prod_{j=1}^{k-2}(n-2(1+j)+1)}{(2k-1)!/2^{k-1}}z^{2k-1}\right],
\end{align}
where the series (\ref{eq:s1}) and (\ref{eq:s2}) have as general terms $c_{2k}$ (\ref{eq:coeffPair}) and $c_{2k-1}$ (\ref{eq:coeffImpair}), respectively. We apply the D'Alembert ratio test and find
\be\label{eq:ConvergenceNew}
\lim_{k\rightarrow\infty}\left|\frac{c_{2(k+1)}}{c_{2k}}\right| = \lim_{k\rightarrow\infty}\left|\frac{c_{2(k+1)-1}}{c_{2k-1}}\right| = 0 <1.
\ee
We conclude that the series (\ref{eq:s1}) and (\ref{eq:s2}) converge. Moreover, these series converge absolutely,  therefore the series $\beta_n(z)$ (\ref{eq:Sol1}) converges.
\\~\\
Case 2: root $\sigma_2 = 1$. 
Consider the series $\nu_n(z)$ (\ref{eq:SerieRacine2}) associated with the root $\sigma_2$ of the indicial equation (\ref{eq:EQDet}). Based on the above reasoning, we obtain
\begin{align}\label{eq:Sol2}
&\nu_n(z) = z-\frac{1}{3}(n-5)z^3+\sum_{k=3}^\infty \left[ (-1)^{k+1} \frac{(n-((2(k-1))^2+1))\prod_{j=1}^{k-2}(n-2(1+j)+1)}{(2k-1)!/2^{k-1}}z^{2k-1}\right],
\end{align}
which is a convergent series, by (\ref{eq:ConvergenceNew}). The expansion of the series $\nu_n(z)$ (\ref{eq:Sol2}) is finite for all values of $n$ which correspond to a root of the polynomial $q_k(n)$ (see the sequence $\lambda_q(k)$ in Table \ref{tab:2}). The coefficients of the series $\nu_n(z)$ (\ref{eq:Sol2}) correspond to the odd coefficients of the series $\beta_n(z)$ (\ref{eq:Sol1}). For this reason, we will now define a notation that will be useful in what follows
\begin{align}
\mu_n(z):&=1-nz^2+\frac{n(n-10)}{6}z^4\label{eq:mu}+\sum_{k=3}^\infty \left[(-1)^k \frac{n(n-((2k-1)^2+1))\prod_{j=1}^{k-2}(n-2(1+j))}{(2k)!/2^{k}}z^{2k}\right],
\end{align}
so that the series $\beta_n(z)$ (\ref{eq:Sol1}) may be rearranged as $\beta_n(z) = \mu_n(z) + \nu_n(z)$.
\begin{remark}
The series $\mu_n(z)$ (\ref{eq:mu}) converges as well, based on relation (\ref{eq:ConvergenceNew}).
\end{remark}
\begin{proposition}\label{th:GenSol}
The series $\beta_n(z)$ (\ref{eq:Sol1}) is a non-polynomial solution of the complex $X_2^{(1)}$-Hermite ODE (\ref{eq:EDOXComplexe}) for all $n\in\mathbb{N}$.
\end{proposition}
\begin{proof}
See Appendix \ref{app:1}.
\end{proof}
\begin{proposition}\label{th:GenSol1}
The series $\mu_{n}(z)$ (\ref{eq:mu}) is a polynomial solution of the complex $X_2^{(1)}$-Hermite ODE (\ref{eq:EDOXComplexe}) for all $n\in2\mathbb{N}\backslash\{2\}$, while the series $\nu_{n}(z)$ (\ref{eq:Sol2}) is a polynomial solution of the complex $X_2^{(1)}$-Hermite ODE (\ref{eq:EDOXComplexe}) for all $n\in(2\mathbb{N}-1)\backslash\{1\}$.
\end{proposition}
\begin{proof}
See Appendix \ref{app:2}.
\end{proof}

We establish the linear dependence relation between $\beta_n(z)$ and $\hat{H}_n(z)$, as well as the linear dependence relation between $\mu_n(z)$ and $\hat{H}_n(z)$ and between $\nu_n(z)$ and $\hat{H}_n(z)$. They are
\begin{align}\label{eq:WronskienBeta}
Wr\left(\hat{H}_n,\beta_n\right)(n;z) &=  \phi_1(n) e^{z^2}(1+2z^2)^2 = \phi_1(n) \frac{16}{W_{(1,1)}(z)},\\\label{eq:Wronskienmu}
Wr\left(\hat{H}_n,\mu_n\right)(n;z) &= 
\left\{\begin{matrix}
\phi_2(n)e^{z^2}(1+2z^2)^2,\;\;\; n\in (2\mathbb{N}-1)\backslash\{1\}\\
0, \qquad\qquad\qquad\quad n\in (2\mathbb{N})\backslash\{2\}\\
\end{matrix}\right.,
\\\label{eq:WronskienHChapeauH2}
Wr\left(\hat{H}_n,\nu_n\right)(n;z) &= 
\left\{\begin{matrix}
0, \qquad\qquad\qquad\qquad \;\,n\in (2\mathbb{N}-1)\backslash\{1\}\\
\phi_3(n)e^{z^2}(1+2z^2)^2,\;\; n\in 2\mathbb{N}\backslash\{2\}\quad\quad\;
\end{matrix}\right.,
\end{align}
where $|\phi_k|:\mathbb{N}\backslash\{1,2\}\rightarrow\mathbb{N}\backslash\{0\}$ is a strictly increasing function, $k = 1, 2, 3$.

The numerators of the coefficients $c_{2k}$ (\ref{eq:coeffPair}) and $c_{2k-1}$ (\ref{eq:coeffImpair}) have no factor in common, therefore the solution $\beta_n(z)$ (\ref{eq:Sol1}) is non-polynomial for all values of $n$.

Consider the solution $\nu_n(z)$ (\ref{eq:Sol2}) and let us define
\begin{align}
r_1(k):&=(2(k-1))^2+1,\qquad k\geq2,\\
r_2(j):&=2(1+j)-1,\qquad\quad\; 1\leq j\leq k-2.
\end{align}
Then the solution $\nu_n(z)$ (\ref{eq:Sol2}) may be written as
\begin{align}\label{eq_Sol2Prime}
&\nu_n(z) = z-\frac{1}{3}(n-r_1(2))z^3+\sum_{k=3}^\infty \left[ (-1)^{k+1} \frac{(n-r_1(k))\prod_{j=1}^{k-2}(n-r_2(j))}{(2k-1)!/2^{k-1}}z^{2k-1}\right].
\end{align}
The roots of the polynomials (\ref{eq:qk})
\begin{align}\label{eq:PolTemp}
&q_k(n) = (n-r_1(k)),\qquad \qquad\quad\qquad\;\, k=2,\\
 &q_k(n)=(n-r_1(k))\prod_{j=1}^{k-2}(n-r_2(j))\quad\;\, k\geq 3,
\end{align}
are odd positive integers as shown in Table \ref{tab:2}
\begin{align}\label{eq:Seq1}
r_1(k)&\in\{5, 17, 37, 65, 101, ...\},\\\label{eq:Seq2}
r_2(j)&\in \{3, 5, 7, ..., 2k-3\}.
\end{align}
Therefore, the solution $\nu_n(z)$ (\ref{eq:Sol2}) is non-polynomial for all $n\in2\mathbb{N}\cup\{1\}$, and polynomial for all odd values of $n$ except $n=1$. The first polynomial cases of the solution $\nu_n(z)$ (\ref{eq:Sol2}) are presented in Table \ref{tab:3}.
\begin{table}[H]
\caption{First polynomial cases of the solution $\nu_n(z)$}
\label{tab:3}       
\centering
\begin{tabular}{lll}
\hline\noalign{\smallskip}
$\quad l$ & $2l-1$ & $\qquad\qquad\qquad\qquad\qquad \nu_{2l-1}(z)$  \\
\noalign{\smallskip}\hline\noalign{\smallskip}
$\quad2$     &  $\quad3$  & $z+\frac{2}{3}z^3$   \\\\
$\quad3$& $\quad5$   & $z+\mathbf{0}\cdot z^3-\frac{4}{5}z^5$     \\\\
$\quad4$&  $\quad7$  & $z-\frac{2}{3}z^3-\frac{4}{3}z^5+\frac{8}{21}z^7$     \\
$\quad\vdots$&    $\quad\;\vdots$  &  $\vdots$     \\
$\quad8$&    $\quad15$  &  $z-\frac{10}{3}z^3-\frac{4}{5}z^5+\frac{88}{21}z^7-\frac{400}{89}z^9+\frac{1376}{3465}z^{11}-\frac{64}{2079}z^{13}+\frac{128}{155925}z^{15}$     \\\\
$\quad9$&    $\quad17$  &  $z-4z^3+\mathbf{0}\cdot z^5+\frac{16}{3}z^7-\cdots-\frac{256}{2297295}z^{17}$     \\\\
$\quad10$&    $\quad19$  &  $z-\frac{14}{3}z^3+\frac{16}{15} z^5+\frac{32}{5}z^7-\cdots+\frac{512}{38513475}z^{19}$     \\
$\quad\vdots$&    $\quad\;\vdots$  &  $\vdots$     \\
$\quad18$&    $\quad35$  &  $z-10z^3+\frac{96}{5}z^5+ \frac{64}{21}z^7-\frac{320}{9}z^9+\cdots +\frac{131072}{6716457438687871875}z^{35}$     \\\\
$\quad19$&    $\quad37$  &  $z-\frac{32}{3}z^3+\frac{68}{3}z^5+\mathbf{0}\cdot z^7-\frac{1088}{27}z^9+\cdots -\frac{262144}{234308415218225473125}z^{37}$     \\
\noalign{\smallskip}\hline
\end{tabular}
\end{table}
Table \ref{tab:3} shows that for the values $n=2l-1\in\{5, 17, 37, ...\}$, one term is missing in the polynomial. As an example, if $n=5$, the third-order term $z^3$ is absent. These values of $n$ correspond to the roots $\lambda_q(k)$ in bold character from Table \ref{tab:2}. This is due to the fact that in these cases, the degree of the polynomial solution $\nu_{2l-1}(z)$ (\ref{eq:Sol2}) corresponds to a root of the polynomials $q_k(n)$ (\ref{eq:PolTemp}), namely a value of the sequence (\ref{eq:Seq1}), \textit{i.e.}
\be
n\in\{r_1(k)\;|\;k\geq2\}\subset \lambda_q(k),
\ee
where $(2k-1)$ is the degree of the missing term.

The coefficient of the first-order term in the polynomials from Table \ref{tab:3} is normalized, because we made the arbitrary choice $\tilde{c}_1 = 1$ during the construction of the generalized series $\nu_{n}(z)$ (\ref{eq:SerieRacine2}). We notice that for each $n\in\{2l-1\;|\;l\geq2\}$, there exists a proportionality constant that depends on $l$, so that
\be
\hat{H}_{2l-1}(z)=M_1(l)\cdot \nu_{2l-1}(z),\qquad l = 2, 3, ...
\ee
As an example, for $l=2$, we have that
\be\label{eq:example}
\hat{H}_3(z) = \frac{(-1)^{(3+1)/2}3!2^{(3+1)/2}}{p_{(1,1)}(3)} \nu_3(z) = 12 \nu_3(z).
\ee
Moreover, equation (\ref{eq:WronskienHChapeauH2}) means that when $n = 2l-1$ for some positive integer $l\geq2$, the series $\nu_{n}(z)$ (\ref{eq:Sol2}) is equal to the polynomial $\hat{H}_{n}(z)$ (\ref{eq:HChapeau}), up to a constant, and we therefore see a part of the $X_2^{(1)}$-Hermite polynomials (corresponding to odd degrees $2l-1\geq3$) arising from the construction of the solution of the ODE (\ref{eq:EDOXComplexe}), using the method of generalized series.
\begin{theorem}\label{th:propHChapeauH2}
The solutions $\hat{H}_{n}(z)$ (\ref{eq:HChapeau}) and $\nu_{n}(z)$ (\ref{eq:Sol2}) of equation (\ref{eq:EDOXComplexe}) follow the proportionality relation
\be
\hat{H}_n(z)  = M_1(n)\nu_n(z),
\ee
for all $n = 2l-1,\; l\geq2$, where
\be\label{eq:M1}
M_1(3) = 12,\quad\qquad  M_1(n) = \frac{(-1)^{(n+1)/2}n!2^{(n+1)/2}}{p_{(1,1)}(n)\prod_{j=1}^{(n-3)/2}(n-2(1+j)+1)},\quad l\geq3,
\ee
where $p_{(1,1)}(n)$ is the polynomial (\ref{eq:polyn_1}) associated with the fixed partition $\lambda = (1)$.
\end{theorem}
\begin{proof}
Let $n = 2l-1,\; l\geq2$, and let $\hat{H}_n(z)$ and $\nu_n(z)$ be the functions defined by (\ref{eq:HChapeau}) and (\ref{eq:Sol2}), respectively. From equations (\ref{eq:Degre2014}) and (\ref{eq:corol_rec}), we know that $\hat{H}_n(z)$ is a polynomial of degree $n$, and therefore equation (\ref{eq:WronskienHChapeauH2}) leads to the conclusion that $\nu_n(z)$ is also a polynomial of degree $n$. The missing term in the solution $\nu_n(z)$ illustrated in Table \ref{tab:3} corresponds to a power of $z$ that is always strictly smaller than $n$, because if there is a missing term, then $n = (2(k-1))^2+1$ for some $k\geq2$.  Since $n = (2(k-1))^2+1>(2k-1)$ for all $k\geq2$, we conclude that the missing term is always associated with a power of $z$ smaller than $n$. The contrary would lead to a contradiction because $\nu_n(z)$ is a polynomial of degree $n$.

Let $\tilde{c}_n$ be the coefficient of the term $z^n$ in the polynomial $\nu_n(z)$. From equation (\ref{eq:Sol2}), we have that for $k\geq3$, the coefficients are defined by the rational expression
\be\label{eq:rational}
\tilde{c}_k(n) = (-1)^{k+1} \frac{(n-((2(k-1))^2+1))\prod_{j=1}^{k-2}(n-2(1+j)+1)}{(2k-1)!/2^{k-1}}.
\ee
The highest-order term $z^n$ is of degree $n = 2k-1$, which implies that $k = (n+1)/2$. Eliminating $k$ in equation (\ref{eq:rational}), we obtain
\be\label{eq:cTilde}
\tilde{c}_n = \frac{(-1)^{(n+1)/2}p_{(1,1)}(n)2^{(n-1)/2}\prod_{j=1}^{(n-3)/2}(n-2(1+j)+1)}{n!}.
\ee
Let $\hat{c}_n$ be the coefficient of the term $z^n$ in the polynomial $\hat{H}_n(z)$. Then, by relation (\ref{eq:HChapeau}), we have that
\be\label{eq:cHat}
\hat{c}_n = 2^n.
\ee
Evaluating the ratio of the coefficients (\ref{eq:cTilde}) and (\ref{eq:cHat}), we get
\be\label{eq:tempM}
M_1(n) = \frac{\hat{c}_n}{\tilde{c}_n} = \frac{(-1)^{(n+1)/2}n!2^{(n+1)/2}}{p_{(1,1)}(n)\prod_{j=1}^{(n-3)/2}(n-2(1+j)+1)}.
\ee
The only zeros of the denominator of $M_1(n)$ are the zeros of $p_{(1,1)}(n)$, namely the gap sequence $\{1,2\}$, because
\be
\left\{\prod_{j=1}^{(n-3)/2}(n-r_2(j))\;:\;n = 3, 5, 7, ...\right\} = \{1, 2, 2\cdot4,2\cdot4\cdot6, ...\},
\ee
which means that $M_1(n)$ is not defined on the gap sequence. We complete the proof by verifying the case when $k=2$, \textit{i.e.} $\hat{H}_3(z)  = M_1(3)\nu_3(z)$ holds. Indeed, $M_1(3) = 12$, which corresponds to the example given in (\ref{eq:example}).
\end{proof}

From Theorem \ref{th:propHChapeauH2}, we find that the function $\phi_2(n)$ in the Wronskian (\ref{eq:Wronskienmu}) corresponds to the additive inverse of the function $M_1(n)$ (\ref{eq:M1})
\be\label{eq:WronskienmuPrime}
Wr\left(\hat{H}_n,\mu_n\right)(n;z) = -M_1(n)e^{z^2}(1+2z^2)^2
\left\{\begin{matrix}
1,\qquad n\in (2\mathbb{N}-1)\backslash\{1\}\\
0, \qquad n\in (2\mathbb{N})\backslash\{2\}\;\;\,\quad\\
\end{matrix}\right..
\ee
Moreover, Theorem \ref{th:propHChapeauH2} sheds light on the fact that the Hermite XOPs $\hat{H}_{2l-1}$ arise from the \textit{odd part} of the series $\beta_n(z)$ (\ref{eq:Sol1}). This motivates the search for the Hermite XOPs $\hat{H}_{2l}$ in the \textit{even part} of the series $\beta_n(z)$ (\ref{eq:Sol1}).

Consider the series $\mu_n(z)$ (\ref{eq:mu}) and let us define
\begin{align}
r_3(k):&=(2k-1)^2+1,\qquad k\geq2,\\
r_4(j):&=2(1+j),\qquad\qquad\; 1\leq j\leq k-2.
\end{align}
Then the series $\mu_n(z)$ (\ref{eq:mu}) may be written as
\begin{align}\label{eq_muPrime}
&\mu_n(z) = 1-nz^2 +\frac{n(n-r_3(2))}{6}z^4+\sum_{k=3}^\infty \left[ (-1)^{k} \frac{n(n-r_3(k))\prod_{j=1}^{k-2}(n-r_4(j))}{(2k)!/2^{k}}z^{2k}\right].
\end{align}
The roots of the polynomials (\ref{eq:pk})
\begin{align}\nonumber
p_k(n) &= (n-r_3(k)),\qquad\qquad \qquad\qquad\;\;\;\; k = 2,\\\label{eq:PolTempmu}
p_k(n) &= n(n-r_3(k))\prod_{j=1}^{k-2}(n-r_4(j)),\qquad k \geq3,
\end{align}
are even positive integers as shown in Table \ref{tab:2}
\begin{align}\label{eq:Seq1mu}
r_3(k)&\in\{10, 26, 50, 82, 122, ...\},\\\label{eq:Seq2mu}
r_4(j)&\in \{4, 6, 8, ..., 2k-2\}.
\end{align}
Therefore, the series $\mu_n(z)$ (\ref{eq:mu}) is non-polynomial for all $n\in(2\mathbb{N}-1)\cup\{2\}$, and is polynomial for all even values of $n$ except $n=2$. The first polynomial cases of the series $\mu_n(z)$ (\ref{eq:mu}) are presented in Table \ref{tab:4}.
\begin{table}[H]
\caption{First polynomial cases of the series $\mu_n(z)$}
\label{tab:4}       
\centering
\begin{tabular}{lll}
\hline\noalign{\smallskip}
$\quad l$ & $\quad2l$ & $\quad\qquad\qquad\qquad\qquad \mu_{2l}(z)$  \\
\noalign{\smallskip}\hline\noalign{\smallskip}
$\quad0$     &  $\quad0$  & $1$   \\\\
$\quad2$& $\quad4$   & $1-4z^2-4z^4$     \\
$\quad\vdots$&    $\quad\;\vdots$  &  $\vdots$     \\
$\quad4$&    $\quad8$  &  $1-8z^2-\frac{8}{3}z^4+\frac{32}{5}z^6-\frac{16}{15}z^8$     \\\\
$\quad5$&    $\quad10$  &  $1-10z^2+\mathbf{0}\cdot z^4+\frac{32}{3}z^6-\frac{80}{21}z^8+\frac{32}{105}z^{10}$     \\\\
$\quad6$&    $\quad12$  &  $1-12z^2+4 z^4+\frac{224}{15}z^6-\cdots-\frac{64}{945}z^{12}$     \\
$\quad\vdots$&    $\quad\;\vdots$  &  $\vdots$     \\
$\quad12$&    $\quad24$  &  $1-24z^2+56z^4+\cdots-\frac{4096}{13749310575}z^{24}$     \\\\
$\quad13$&    $\quad26$  &  $1-26z^2+\frac{208}{3}z^4+\mathbf{0}\cdot z^6-\cdots+\frac{8192}{316234143225}z^{26}$     \\
\noalign{\smallskip}\hline
\end{tabular}
\end{table}
Table \ref{tab:4} shows that for the values $n=2l\in\{10, 26, 50, ...\}$, one term is missing in the polynomial. As an example, if $n=10$, the fourth-order term $z^4$ is absent. These values of $n$ correspond to the roots $\lambda_p(k)$ in bold character from Table \ref{tab:2}. This is due to the fact that in these cases, the degree of the polynomial $\mu_n(z)$ (\ref{eq:mu}) corresponds to a root of the polynomials $p_k(n)$ (\ref{eq:PolTempmu}), namely a value of the sequence (\ref{eq:Seq1mu}), \textit{i.e.}
\be
n\in\{r_3(k)\;|\;k\geq2\}\subset \lambda_p(k),
\ee
where $(2k)$ is the degree of the missing term.

The constant term in the polynomials from Table \ref{tab:4} is normalized, because we made the arbitrary choice $c_1 = 1$ during the construction of the generalized series $\beta_n(z)$ (\ref{eq:SerieRacine1}). We notice that for each $n\in\{2l\;|\;l\geq2\}$, there exists a proportionality constant that depends on $l$, so that
\be
\hat{H}_{2l}(z)=M_2(l)\cdot \mu_{2l}(z),\qquad l = 2, 3, ...
\ee
\begin{theorem}\label{th:propHChapeaumu}
The functions $\hat{H}_n(z)$ (\ref{eq:HChapeau}) and $\mu_n(z)$ (\ref{eq:mu}) follow the proportionality relation
\be
\hat{H}_n(z)  = M_2(n)\mu_n(z),
\ee
for all $n = 2l,\; l\in\{0, 2, 3, 4, ...\}$, where
\begin{align}\label{eq:M2}
&M_2(0):=1, \quad\qquad M_2(n) = \frac{(-1)^{(n+2)/2}n!2^{n/2}}{n \cdot p_{(1,1)}(n)\prod_{j=1}^{(n-4)/2}(n-2(1+j))},\quad n\geq1.
\end{align}
\end{theorem}
\begin{proof}
Let $n = 2l,\; l\geq2$, and let $\hat{H}_n(z)$ and $\mu_n(z)$ be the functions defined by (\ref{eq:HChapeau}) and (\ref{eq:mu}), respectively. Then $\mu_n(z)$ is a polynomial of degree $n$. The  missing term in the polynomial $\mu_n(z)$ presented in Table \ref{tab:4} corresponds to a power of $z$ that is always strictly smaller than $n$, because if there is a missing term, then $n = (2k-1)^2+1$ for some $k\geq2$.  Since $n = (2k-1)^2+1>2k$ for all $k\geq2$, we conclude that the missing term is always associated with a power of $z$ smaller than $n$.

Let $c_n$ be the coefficient of the term $z^n$ in the polynomial $\mu_n(z)$. From equation (\ref{eq:mu}), we have that for $k\geq3$, the coefficients are defined by the rational expression
\be\label{eq:rationalmu}
c_k(n) = (-1)^{k} \frac{n(n-((2k-1)^2+1))\prod_{j=1}^{k-2}(n-2(1+j))}{(2k)!/2^{k}}.
\ee
The highest-order term $z^n$ is of degree $n = 2k$, which means that $k = n/2$. Eliminating $k$ in equation (\ref{eq:rationalmu}), we obtain
\be\label{eq:cmu}
c_n = \frac{(-1)^{(n+1)/2}p_{(1,1)}(n)2^{(n-1)/2}\prod_{j=1}^{(n-3)/2}(n-2(1+j)+1)}{n!}.
\ee
Let $\check{c}_n$ be the coefficient of the term $z^n$ in the polynomial $\hat{H}_n(z)$. Then, by relation (\ref{eq:HChapeau}), we have that
\be\label{eq:cHatmu}
\check{c}_n = 2^{2n}.
\ee
Evaluating the ratio of the coefficients (\ref{eq:cmu}) and (\ref{eq:cHatmu}), we get
\be\label{eq:tempMmu}
M_2(n) = \frac{\check{c}_n}{c_n} = \frac{(-1)^{(n+2)/2}n!2^{n/2}}{n \cdot p_{(1,1)}(n)\prod_{j=1}^{(n-4)/2}(n-2(1+j))}.
\ee
The only zeros of the denominator of $M_2(n)$ are the zeros of $p_{(1,1)}(n)$ and zero itself, namely the gap sequence $\{1,2\}$ and zero, because
\be
\left\{\prod_{j=1}^{(n-4)/2}(n-r_4(j))\;:\;n = 4, 6, 8, ...\right\} = \{1, 3, 3\cdot5,3\cdot5\cdot7, ...\},
\ee
which means that $M_2(n)$ is not defined on the gap sequence, neither at $n=0$, which is why we define this case separately in equations (\ref{eq:M2}) (Remark \ref{rem:6} contains another formulation of the constant $M_2(n)$).

We complete the proof by verifying the case $k=2$. Indeed, $\hat{H}_4(z)  = M_2(4)\mu_4(z)$, where $M_2(4)=-4$.
\end{proof}
\begin{remark}\label{rem:6}
Since $n$ is even, $M_2(n)$ can be equivalently expressed in terms of the Euler gamma function \cite{Nikiforov1988}
\be
M_2(n) = (-1)^{(n+2)/2}2^{n-1}\pi^{-1/2}\Gamma\left(\frac{n-1}{2}\right),\qquad \forall n\in 2\mathbb{N}\backslash\{2\}.
\ee
\end{remark}

From Theorem \ref{th:propHChapeaumu}, we find that the function $\phi_3(n)$ in the Wronskian (\ref{eq:WronskienHChapeauH2}) corresponds to the function $M_2(n)$ (\ref{eq:M2})
\be\label{eq:WronskienHChapeauH2Prime}
Wr\left(\hat{H}_n,\nu_n\right)(n;z) = M_2(n)e^{z^2}(1+2z^2)^2\cdot 
\left\{\begin{matrix}
0, \qquad n\in (2\mathbb{N}-1)\backslash\{1\}\\
1,\qquad n\in 2\mathbb{N}\backslash\{2\}\qquad\;\;
\end{matrix}\right..
\ee
From Theorems \ref{th:propHChapeauH2} and \ref{th:propHChapeaumu}, we find that the function $\phi_1(n)$ appearing in the Wronskian (\ref{eq:WronskienBeta}) corresponds to the additive inverse of the function $M_1(n)$ (\ref{eq:M1}) when $n$ is odd, and to the function $M_2(n)$ (\ref{eq:M2}) when $n$ is even
\be\label{eq:WronskienBetaPrime}
Wr\left(\hat{H}_n,\beta_n\right)(n;z) =   e^{z^2}(1+z^2)^2 \cdot
\left\{\begin{matrix}
-M_1(n), \qquad n\in (2\mathbb{N}-1)\backslash\{1\}\\
\;M_2(n), \qquad n\in 2\mathbb{N}\backslash\{2\}\qquad
\end{matrix}\right..     
\ee
\begin{corollary}\label{cor:Prop2019H2}
The functions $H^{(1)}_n(z)$ (\ref{eq:HermiteExceptFixe}), $\mu_n(z)$ (\ref{eq:mu}) and $\nu_n(z)$ (\ref{eq:Sol2}) follow the proportionality relations
\be
H^{(1)}_n(z) = \frac{(-1)^{(n+1)/2}8n!2^{(n+1)/2}}{\prod_{j=1}^{(n-3)/2}(n-2(1+j)+1)}\nu_n(z)
\ee
for all $n = 2l-1,\; l\geq2$,
\be
H^{(1)}_n(z) = \frac{(-1)^{(n+2)/2}8n!2^{n/2}}{n \prod_{j=1}^{(n-4)/2}(n-2(1+j))}\mu_n(z)
\ee
for all $n = 2l,\; l\geq2,$
\be
H^{(1)}_0(z) = 16\mu_0(z).
\ee
\end{corollary}
\begin{proof}
The proof is straightforward, making use of Theorems \ref{cor:Poynome2019}-\ref{th:propHChapeaumu}. 
\end{proof}
\begin{corollary}\label{cor:OrthogonalityH2}
The functions $\nu_n(z)$ (\ref{eq:Sol2}) and $\mu_n(z)$ (\ref{eq:mu}) follow the orthogonality relations
\be
\int_{-\infty}^\infty \nu_m(x)\nu_n(x)\frac{e^{-x^2}}{(1+2x^2)^2}\;dx = \delta_{m,n}\frac{\sqrt{\pi}p_{(1,1)}(n)\prod_{j=1}^{(n-3)/2}(n-2(1+j)+1)^2}{2\cdot n!}
\ee
for all $n = 2l-1,\; l\geq2$,
\be
\int_{-\infty}^\infty \mu_m(x)\mu_n(x)\frac{e^{-x^2}}{(1+2x^2)^2}\;dx = \delta_{m,n}\frac{\sqrt{\pi}\,n\cdot p_{(1,1)}(n)\prod_{j=1}^{(n-4)/2}(n-2(1+j))^2}{ (n-1)!}
\ee
for all $n = 2l,\; l\geq2$,
\be
\int_{-\infty}^\infty \left(\mu_0(x)\right)^2\frac{e^{-x^2}}{(1+2x^2)^2}\;dx = \frac{\sqrt{\pi}}{2}.
\ee
\end{corollary}
\begin{proof}
If $n = 2l-1,\; l\geq2$, then by virtue of Theorem \ref{th:propHChapeauH2}, the orthogonality relation (\ref{eq:orthogPolynome2019}) for $\hat{H}_n(x)$ becomes
\begin{align}\label{eq:tempOrthog3}
\int_{-\infty}^{+\infty}\nu_{m}(x)\nu_{n}(x)\frac{e^{-x^2}}{(1+2x^2)^2}dx &= \delta_{m,n}M_1^{-2}(n)\frac{\sqrt{\pi}2^{n}n!}{(n-1)(n-2)}\\\nonumber
&=\delta_{m,n}\frac{\sqrt{\pi}p_{(1,1)}(n)\prod_{j=1}^{(n-3)/2}(n-2(1+j)+1)^2}{2\cdot n!}.
\end{align}
The case $n=0$ is easily verified. If $n = 2l,\; l\geq2$, then by virtue of Theorem \ref{th:propHChapeaumu}, the orthogonality relation (\ref{eq:orthogPolynome2019}) for $\hat{H}_n(x)$ becomes
\begin{align}\label{eq:tempOrthog3mu}
\int_{-\infty}^{+\infty}\mu_{m}(x)\mu_{n}(x)\frac{e^{-x^2}}{(1+2x^2)^2}dx &= \delta_{m,n}M_2^{-2}(n)\frac{\sqrt{\pi}2^{n}n!}{(n-1)(n-2)}\\\nonumber
&=\delta_{m,n}\frac{\sqrt{\pi}\,n\cdot p_{(1,1)}(n)\prod_{j=1}^{(n-4)/2}(n-2(1+j))^2}{ (n-1)!}.
\end{align}
\end{proof}
\begin{remark}
On the gap sequence, the functions $\hat{H}_{1}(z)$ and $\hat{H}_{2}(z)$ are  defined from the extension of the parameter of classical Hermite polynomials to negative integers, as in equations (\ref{eq:HermiteNegInt}), leading to non-polynomial solutions of the ODE (\ref{eq:EDOXComplexe}). Theorems \ref{th:propHChapeauH2} and \ref{th:propHChapeaumu} state that there exists a proportionality constant between the polynomials $\hat{H}_{2l}(z)$ (\ref{eq:HChapeau}) and $\mu_{2l}(z)$ (\ref{eq:mu}), where $l\in\{0, 2, 3, ...\}$ and between the polynomials $\hat{H}_{2l-1}(z)$ (\ref{eq:HChapeau}) and $\nu_{2l-1}(z)$ (\ref{eq:Sol2}), where $l\in\{2, 3, 5, ...\}$. The same phenomenon holds between $\hat{H}_{2}(z)$ and $\mu_{2}(z)$ and between $\hat{H}_{1}(z)$ and $\nu_{1}(z)$, namely on the gap sequence, which indicates that the extension of classical Hermite polynomials to negative integers arises naturally in the construction of the solution of the ODE (\ref{eq:EDOXComplexe}).
\end{remark}
We showed in the present section that the $X_2^{(1)}$-Hermite polynomials of even degree may be expressed as the \textit{even part} of the series $\beta_{n}(z)$ (\ref{eq:Sol1}) and that the $X_2^{(1)}$-Hermite polynomials of odd degree may be expressed as the \textit{odd part} of the series $\beta_{n}(z)$ (\ref{eq:Sol1}). That suggests that the general solution of the complex $X_2^{(1)}$-Hermite ODE (\ref{eq:EDOXComplexe}) can be expressed as a linear combination of two separate functions, where the first function would include the $X_2^{(1)}$-Hermite polynomials together with the non-polynomial cases $n=1,2$ (the gap sequence), and where the second function would include non-polynomial solutions.

Let $\hat{H}_{1}(z)$ and $\hat{H}_{2}(z)$ be defined by (\ref{eq:HermiteNegInt}). Then $\{\hat{H}_{n}(z)\}_{n=0}^\infty$ is a countable sequence of functions which includes the $X_2^{(1)}$-Hermite polynomials, up to a constant, and two non-polynomial solutions, namely for $n=1,2$ (the gap sequence). Consider the proportionality constants $M_1(n)$ and $M_2(n)$ from Theorems \ref{th:propHChapeauH2} and \ref{th:propHChapeaumu}, respectively, and let us define
\be\label{eq:M3}
M_3(n) := \left\{\begin{matrix}
-2, \qquad\;\; n=1\quad\;\;\;\;\;\;\,\\
2,\qquad\;\; n=2\quad\;\;\;\,\\
-M_1^{-1}(n),  \quad n \in 2\mathbb{N}-1\backslash\{1\}\\
\;\;M_2^{-1}(n), \quad n \in 2\mathbb{N}\backslash\{2\}\quad\;\;
\end{matrix}\right.,
\ee
and
\be\label{eq:alpha}
\alpha_n(z) := M_3(n)\hat{H}_{n}(z), \qquad n\in \mathbb{N}.
\ee
We introduced the constant $M_3(n)$ in the definition (\ref{eq:alpha}) in order to obtain simplified and normalized Wronskians in the various calculations involving the functions $\alpha_n(z), \beta_n(z), \mu_n(z)$, and $\nu_n(z)$.

Under the above assumptions, we have the following theorem.
\begin{theorem} (\textbf{Main result}) \label{th:Main}
The general solution of the complex $X_2^{(1)}$-Hermite ODE (\ref{eq:EDOXComplexe}) is given by
\be\label{eq:GenSolDecomposed}
\omega_n(z) = a_1\alpha_n(z)+a_2\beta_n(z), \qquad n\in\mathbb{N},\;\;\: a_1, a_2 \in\mathbb{C},
\ee
where the functions $\alpha_n(z)$ and $\beta_n(z)$ are given by (\ref{eq:alpha}) and (\ref{eq:Sol1}), respectively, and are linearly independent. These functions have the following properties:
\\~\\
(i) The elements of the countable sequence $\{\alpha_{n}(z)\}_{n\in\mathbb{N}\backslash\{1,2\}}$ correspond to the complex exceptional Hermite polynomials with the partition $\lambda = (1)$.

The functions $\{\alpha_{n}(z)\}_{n\in\{1,2\}}$ correspond to the non-polynomial solutions (\ref{eq:HermiteNegInt}) which complete the gap for $n=1,2$.
\\~\\
(ii) The elements of the countable sequence $\{\beta_{n}(z)\}_{n\in\mathbb{N}}$ are non-polynomial functions.
\end{theorem}
\begin{proof}
\textit{(i)} Consider the definition of the function $\alpha_n(z)$ (\ref{eq:alpha}). By Corollary \ref{cor:HChapeau}, we know that $\hat{H}_n(z)$ is a polynomial solution of the complex $X_2^{(1)}$-Hermite ODE (\ref{eq:EDOXComplexe}) for all $n\in\mathbb{N}\backslash\{1,2\}$, which indicates that $\alpha_n(z)$ is a polynomial solution for these values of $n$. The non-polynomial cases for $n=1,2$ are easily verified by substituting the expressions (\ref{eq:HermiteNegInt}) and their derivatives up to order 2 into the ODE (\ref{eq:EDOXComplexe}). Making use of Theorems \ref{th:propHChapeauH2} and \ref{th:propHChapeaumu}, we obtain the Wronskians
\be\label{eq:WronskienmuComplete}
Wr\left(\alpha_n,\mu_n\right)(z) = e^{z^2}(1+2z^2)^2\cdot
\left\{\begin{matrix}
1,& \qquad\, n\in 2\mathbb{N}-1\\
0, & \;\; n\in 2\mathbb{N}
\end{matrix}\right.,
\ee
\be\label{eq:WronskienHChapeauH2Complete}
Wr\left(\alpha_n,\nu_n\right)(z) = e^{z^2}(1+2z^2)^2\cdot
\left\{\begin{matrix}
0,  & \qquad\, n\in 2\mathbb{N}-1\\
1,&\;\;n\in 2\mathbb{N}
\end{matrix}\right.,
\ee
which show that $\{\hat{H}_{2l}(z),\mu_{2l}(z)\}$ and $\{\hat{H}_{2l+1}(z),\nu_{2l+1}(z)\}$ are linearly dependent sets for all $l\geq0$ (it was shown in Appendix \ref{app:2} that $\{\mu_{2l}(z)\}$ and $\{\nu_{2l+1}(z)\}$ are solutions for all $l\geq0$). By Proposition \ref{th:GenSol}, we know that $\beta_n(z)$ is a non-polynomial solution of the ODE (\ref{eq:EDOXComplexe}) for all $n\in\mathbb{N}$. The Wronskian
\be\label{eq:WronskienBetaComplete}
Wr\left(\alpha_n,\beta_n\right)(z) =   e^{z^2}(1+z^2)^2 \cdot
\left\{\begin{matrix}
1-\sqrt{\pi}/2,&\;\,n=1\qquad\quad\;\;\\
1-2\sqrt{\pi},&\;\,n=2\qquad\quad\;\;\\
1, & n\in \mathbb{N}\backslash\{1,2\}
\end{matrix}\right.,    
\ee
shows that $\alpha_n(z)$ and $\beta_n(z)$ are linearly independent functions. 

\textit{(ii)} The sequence $\{\beta_{n}(z)\}_{n\in\mathbb{N}}$ is composed of non-polynomial functions. This is due to the form of the series $\beta_n(z)$ (\ref{eq:Sol1}), composed of coefficients $c_{2k}(n)$ (\ref{eq:coeffPair}) and $c_{2k-1}(n)$ (\ref{eq:coeffImpair}). They are asociated with even powers of $z$ and odd powers of $z$, respectively and have no root in common (see the roots $\lambda_p(k)$ and $\lambda_q(k)$ in Table \ref{tab:2}).
\end{proof}
\begin{remark}
The function $\omega_n(z)$ (\ref{eq:GenSolDecomposed}) is the analytical general solution of the ODE (\ref{eq:OperatorTlambdaSquared}) for $n\in\mathbb{N}$ (no gap), on the complex plane, for the particular case $\lambda = (1)$. Provided that $\lambda^2$ is an Adler partition, the differential operator $T_{\lambda^2}$ is non-singular on $\mathbb{R}$. However, in the case $\lambda = (1)$ the operator possesses singularities at ${\pm i/\sqrt{2}}\notin\mathbb{R}$. This is due to the fact that the operator $T_\lambda$ (\ref{eq:HermioteXOp}) has singularities corresponding to the zeros of the Wronskian $H_\lambda$ (\ref{eq:wronskiens}).  Indeed, the general solution was built from an Adler partition, leading to a gap sequence which fulfills the hypotheses of the Krein-Adler Theorem \ref{th:1}. Consequently, the ODE (\ref{eq:EDOXComplexe}) arising from these choices has no singularity on $\mathbb{R}$. 

The solution $\alpha_n(z)$ (\ref{eq:alpha}) is non-polynomial on the gap sequence arising from the choice of the partition $\lambda = (1)$. The expression for the coefficients of the series $\beta_n(z)$ (\ref{eq:Sol1}) constructed from the differential  operator $T_{(1,1)}$ provides some clues explaining the existence of a gap in the eigenvalue spectrum of the differential operator.  The coefficients $c_{2k}(n)$ (\ref{eq:coeffPair}) and $c_{2k-1}(n)$ (\ref{eq:coeffImpair}) of the series $\beta_n(z)$ (\ref{eq:Sol1}), associated with even and odd powers of $z$, repectively, are polynomials of the parameter $n$, which possess no root on the gap sequence (see Table \ref{tab:2}), leading to the non-polynomial solutions $\nu_1(z)$ and $\mu_2(z)$ for the values $n=1,2$.
\end{remark}
\section{Minimal surface representation of the general solution of the $X_2^{(1)}$-Hermite ODE}
\label{sec:4}
In this section, we present the geometric representation of the general solution of the $X_2^{(1)}$-Hermite ODE (\ref{eq:EDOXComplexe}), the function $\omega_n(z)$ (\ref{eq:GenSolDecomposed}), under the form of a family of minimal surfaces.

For the purpose of construction of minimal surfaces, we make use of the link between the classical Enneper-Weierstrass formula for the immersion of a minimal surface $F$ in the Euclidean space $\mathbb{E}^3$ and the linear problem for the moving frame 
\be\label{eq:MovingFrame}
\sigma = (\pa F, \opa F, N)^T
\ee
on the surface, where we used the following notation for the holomorphic and antiholomorphic derivatives
\be
\pa = \frac{1}{2}\left(\frac{\pa}{\pa x}-i\frac{\pa}{\pa y}\right),\qquad \opa = \frac{1}{2}\left(\frac{\pa}{\pa x}+i\frac{\pa}{\pa y}\right), \qquad z = x+iy.
\ee
This link, expressed as a second-order linear ODE, allows us to prove  that the general solution (\ref{eq:GenSolDecomposed}) can be represented by minimal surfaces. We calculate the explicit form of the immersion formula and we present a numerical display of these surfaces for different values of the parameter of the complex $X_2^{(1)}$-Hermite ODE (\ref{eq:EDOXComplexe}).
\subsection{Enneper-Weierstrass formula and $\frak{su}(2)$ representation}\label{sec:Enneper}
Consider the Enneper-Weierstrass immersion formula \cite{enneper1868analytisch,weierstrass1866fortsetzung} describing a zero mean curvature surface (denoted by $h=0$) in terms of two locally holomorphic arbitrary functions
\begin{equation}
 \label{eq:F}F(\xi_0,\xi) = \frac{1}{2}\mathbb{R}e\left( \int_{\xi_{0}}^\xi \left( 1 - \chi^2,\; i(1 + \chi^2),\; 2\chi)\right)^T\eta^2 \; dz\right) \in \mathbb{E}^3,
\end{equation}
where $\opa\eta = \opa\chi = 0$. The integration in formula (\ref{eq:F}) is performed on an arbitrary path from the constant $\xi_0\in\mathbb{C}$ to the complex variable $\xi\in\mathbb{C}\backslash\{\xi_0\}$.

Let $\tilde{F}\in\frak{su}(2)\simeq \mathbb{E}^3$ be the quaternionic description of the minimal surface represented by formula (\ref{eq:F}). In order to determine the explicit form of this representation, we identify the Euclidean space $\mathbb{E}^3$ with the imaginary quaternions \cite{Bobenko1994} by the formula
\begin{equation}\label{eq:Desc_Quatern} \tilde{F} = -i\sum_{k = 1}^3F_k\sigma_k \in \mathbb{I}m\mathbb{H}\simeq \frak{su}(2),\qquad Tr(\tilde{F})=0 ,\quad \tilde{F}^\dagger = -\tilde{F},
\end{equation} 
where dagger $\dagger$ denotes the Hermitian conjugate of the considered expression. The matrices $\sigma_k, k = 1,2,3$ are the Pauli matrices, such that $\sigma_k^\dagger = \sigma_k$. The inner product is then
 \be
 \braket{X,Y} = -\frac{1}{2}Tr(XY),\qquad \forall X,Y \in \frak{su}(2).
 \ee
Substituting the components $F_k, k = 1,2,3,$ of the Enneper-Weierstrass representation (\ref{eq:F}) into formula (\ref{eq:Desc_Quatern}), we obtain a matrix formulation of the surface  \cite{Chalifour2019}
\begin{equation}\label{eq:53}
 \tilde{F} = -\frac{i}{2}\left(\begin{array}{cc}  \int_{\xi_{0}}^\xi \chi\eta^2 \;dz + \left(\int_{\xi_{0}}^\xi \chi\eta^2 \;dz \right)^* & \int_{\xi_{0}}^\xi \eta^2 \;dz   -    \left(\int_{\xi_{0}}^\xi \chi^2\eta^2 \;dz \right)^* \\    \\ -\int_{\xi_{0}}^\xi \chi^2\eta^2 \;dz   +    \left(\int_{\xi_{0}}^\xi \eta^2 \;dz \right)^*  & - \int_{\xi_{0}}^\xi \chi\eta^2 \;dz  - \left(\int_{\xi_{0}}^\xi \chi\eta^2 \;dz \right)^* \\ \end{array}\right),
 \end{equation}
where star $*$ denotes the complex conjugate of the considered expression.  The formula (\ref{eq:53}) for $\tilde{F}$ is a quaternionic representation of the surface immersed in the $\frak{su}(2)$ Lie algebra, because $Tr(\tilde{F}) = 0$ and $\tilde{F}^\dagger = -\tilde{F}$.
\subsection{Holomorphic reduction of the linear problem for the moving frame}\label{sec:42}
Making use of the Lie algebra isomorphism $\frak{so}(3) \simeq \frak{su}(2)$, the Gauss-Weingarten equations for the moving frame $\sigma$ (\ref{eq:MovingFrame}) may be written in terms of $2\times2$ complex-valued matrices \cite{Bobenko1994,Bobenko2000}. When the mean curvature vanishes ($h=0$), the wavefunction $\Phi \in SU(2, \mathbb{C})$ satisfies the linear differential equations
\begin{align}\label{eq:15}\partial\Phi = \mathcal{U}\Phi, \qquad \opa\Phi = \mathcal{V}\Phi,\qquad\qquad\qquad\quad\\\nonumber\\\nonumber
\mathcal{U} = \left( \begin{array}{cc}
 \frac{1}{4}\partial u & -Qe^{-\frac{u}{2}} \\
 0 & -\frac{1}{4}\partial u \\
\end{array}\right),\quad 
\mathcal{V} = \left( \begin{array}{cc}
 -\frac{1}{4}\overline{\partial}u & 0 \\
 \overline{Q}e^{-\frac{u}{2}} & \frac{1}{4}\overline{\partial}u\\
\end{array}\right)\in \frak{sl}(2,\mathbb{C}),
\end{align}
where $\mathcal{U}^\dagger = -\mathcal{V}$. The Euclidean metric $\Omega = e^u dzd\bar{z}$ on the surface $\tilde{F}$ is conformal, where $z, \bar{z}$ are local coordinates on $\tilde{F}$ and $Qdz^2$ is the Hopf differential. We apply the gauge transformation $M$ to the wavefunction $\Phi\in SU(2, \mathbb{C})$ of the linear problem (\ref{eq:15}), as proposed in \cite{Doliwa2012} and used afterwards in \cite{Chalifour2019}
\begin{equation}\label{eq:31_0}
\Psi = M\Phi, \qquad \text{ where } \;M =\left( \begin{array}{cc}\frac{|\eta|(1+\chi\overline{\chi})^{1/2}}{\eta\chi}&0\\\\ -\frac{|\eta|}{\eta(1+\chi\overline{\chi})^{1/2}}&\frac{\eta\chi}{|\eta|(1+\chi\overline{\chi})^{1/2}}  \end{array}\right)\in SL(2, \mathbb{C}).
\end{equation}
We obtain
\begin{equation}\label{eq:32}
\partial\Psi = \tilde{\mathcal{U}}(\lambda;z)\Psi, \qquad \opa \Psi=\mathbf{0},
\end{equation}
where
\begin{equation}\label{eq:50}
\tilde{\mathcal{U}}(\lambda;z) = \lambda\eta^2\left( \begin{array}{cc}    \chi & -1\\ \chi^2 & -\chi   \end{array}\right),\quad \tilde{\mathcal{V}} = \mathbf{0} \in\frak{sl}(2,\mathbb{C}).
\end{equation}
The system (\ref{eq:32}) is a reduced linear problem for the holomorphic wavefunction $\Psi(z)$. The potential matrix $\tilde{\mathcal{U}}$ is parametrized by the spectral parameter $\lambda\in \mathbb{C}\backslash\{0\}$, where 
\be\label{eq:tempparamer}
\eta = re^{i\theta}, \quad r >0,\quad \theta\in [0, 2\pi[, \quad \lambda = \eta/\overline{\eta} = e^{2i\theta}.
\ee 
\begin{corollary}\label{cor:new1}
The $SU(2,\mathbb{C})$ wavefunction $\Phi$ of a minimal surface given by (\ref{eq:15}) is gauge equivalent to the wavefunction $\Psi$ of the linear system (\ref{eq:32}) with potential matrices (\ref{eq:50}) together with (\ref{eq:tempparamer}). The gauge is given by (\ref{eq:31_0}). \end{corollary}

In what follows, we consider the linear system (\ref{eq:32}) with a two-component holomorphic vector wavefunction $\tilde{\Psi} = (\psi_1, \psi_2)^T$. Consequently, the linear system (\ref{eq:32}) can be equivalently expressed by the system
\begin{align}\label{eq:52}
&\partial^2\psi_1 - 2\frac{\partial\eta}{\eta}\partial \psi_1 - \lambda\eta^2\partial\chi\psi_1 = 0,\\\label{eq:52_a}
&\psi_2 = \chi \psi_1 - \frac{\partial \psi_1}{\lambda \eta^2}.
\end{align}
\begin{corollary}\label{cor:new}
The identification of the ODE (\ref{eq:52}) with any selected linear second-order ODE leads to the explicit determination of the holomorphic functions $\eta$ and $\chi$ from the Enneper-Weierstrass formula (\ref{eq:F}). Consequently, it is possible to determine a $\frak{su}(2)$-valued minimal surface representation $\tilde{F}$ (\ref{eq:53}) which corresponds to solutions of the selected ODE.
 \end{corollary}
\subsection{Links between the linear problem and the $X_2^{(1)}$-Hermite differential equation}
Let us identify the coefficients of the linear second-order ODE (\ref{eq:52}) with the coefficients (\ref{eq:coeff}) of the complex $X_2^{(1)}$-Hermite ODE (\ref{eq:EDOXComplexe}). We obtain the system
\be\label{eq:Association}
\frac{\partial\eta}{\eta} = z+\frac{4z}{1+2z^2}, \qquad - \lambda\eta^2\partial\chi = 2n.
\ee
This identification signifies that the component $\psi_1(n;z)$ of the holomorphic wavefunction $\tilde{\Psi}$ corresponds to the general solution $\omega_n(z)$ of equation (\ref{eq:EDOXComplexe}). We obtain the explicit form of the arbitrary functions from the Enneper-Weierstrass representation (\ref{eq:F})
\begin{align}\label{eq:etaExceptional}
\eta^2(z)&= \frac{16c_1^2}{W_{(1,1)}(z)},\\\label{eq:chiExceptional}
\chi(n;\lambda;z) 
 &= -\frac{2n}{\lambda c_1^2}\left(c_2 +\frac{\sqrt{\pi}}{4}\mathrm{erf}(z)+8(1+2z^2)W_{(1,1)}(z)\right),
\end{align}
where $W_{(1,1)}(z)$ is the weight function (\ref{eq:Poidslambda1}) and $c_1\in\mathbb{C}\backslash\{0\}$ and $c_2\in\mathbb{C}$ are arbitrary constants. The functions $\eta$ (\ref{eq:etaExceptional}) and $\chi$ (\ref{eq:chiExceptional}) are written in terms of the complex extension of the weight (\ref{eq:Poidslambda1}).
 Substituting the functions $\eta$ (\ref{eq:etaExceptional}) and $\chi$ (\ref{eq:chiExceptional}) into (\ref{eq:50}), we obtain the components of the potential matrix $\tilde{\mathcal{U}}(n; \lambda;z) = (u_{ij})$ in the form
\begin{align}\nonumber
u_{11} &=-u_{22} =  -\frac{32n}{W_{(1,1)}(z)}\left(c_2+\frac{\sqrt{\pi}}{4}\mathrm{erf}(z)+8(1+2z^2)W_{(1,1)}(z)\right),\\
u_{12} &= -\frac{16\lambda c_1^2}{W_{(1,1)}(z)},\\\nonumber
u_{21} &= \frac{64n^2}{\lambda c_1^2W_{(1,1)}(z)}\left(c_2+\frac{\sqrt{\pi}}{4}\mathrm{erf}(z)+8(1+2z^2)W_{(1,1)}(z)\right)^2.
\end{align}
Given the conditions (\ref{eq:52})-(\ref{eq:52_a}) on its components $\psi_1$ and $\psi_2$, the wavefunction takes the form
\be\label{eq:WavefunctionFinal}
\tilde{\Psi}(n; \lambda; z) =  \left(\begin{matrix}\omega_n(z)\\\\
-\frac{2n}{\lambda c_1^2}\left(c_2 +\frac{\sqrt{\pi}}{4}\mathrm{erf}(z)+8(1+2z^2)W_{(1,1)}(z)\right) \omega_n(z) 
- \frac{1}{16\lambda c_1^2}\partial \omega_n(z)W_{(1,1)}(z)\end{matrix}\right),
\ee
where $c_1,\lambda \in\mathbb{C}\backslash\{0\}$ and $c_2\in\mathbb{C}$. Note that the wavefunction $\tilde{\Psi}$ is expressed here in terms of the general solution $\omega_n(z)$ (\ref{eq:GenSolDecomposed}) and its derivative.
\subsection{Minimal surfaces describing the general solution $\omega_n(z)$}
The explicit form of the components of the Enneper-Weierstrass representation (\ref{eq:F}) is obtained by integration of the functions $\eta$ (\ref{eq:etaExceptional}) and $\chi$ (\ref{eq:chiExceptional}). Let us denote
\begin{equation}\label{eq:Integrales}
I_1:= \int_{\xi_{0}}^\xi \eta^2\;dz, \quad\qquad I_2 := \int_{\xi_{0}}^\xi \chi^2\eta^2\;dz, \quad\qquad I_3:=\int_{\xi_{0}}^\xi \chi\eta^2\;dz.
\end{equation}
Then the Enneper-Weierstrass immersion formula (\ref{eq:F}) describing a minimal surface immersed in $\mathbb{E}^3$ becomes
\begin{equation}\label{eq:Integrales2}
 F(n; \lambda;\xi_0, \xi) = \left(\frac{1}{2}\mathbb{R}e\left( I_1 - I_2 \right), -\frac{1}{2}\mathbb{I}m\left( I_1 + I_2 \right), \mathbb{R}e\left(  I_3 \right)\right)^T \;\; \in \mathbb{E}^3,
\end{equation}
where
\begin{align}\label{eq:I1}
I_1 = &c_1^2\left[  \sqrt{\pi} \mathrm{erfi}(z)+e^{z^2}z(2z^2-1) \right]_{\xi_0}^\xi,
\\\nonumber\\
\label{eq:I2}
I_2 = &\frac{4n^2}{\lambda^2 c_1^2}\left[  c_2^2\sqrt{\pi}\mathrm{erfi}(z)+\frac{\sqrt{\pi}}{6}z^2 \mathrm{erf}(z)+ \frac{\sqrt{\pi}}{4}z \mathrm{erf}(z)+\frac{\sqrt{\pi}}{8} \mathrm{erf}(z)-\frac{c_2}{2}z^2\left.\right._{2}F_{2}(1,1;-1/2,2;z^2)\right.\\\nonumber
&\quad\left.+c_2\sqrt{\pi}z^2\left.\right._{2}F_{2}(1,1;1/2,2;z^2)+\frac{c_2\sqrt{\pi}}{2}z^2 \left.\right._{2}F_{2}(1,1;3/2,2;z^2)-\frac{c_2\sqrt{\pi}}{2} z^4+\frac{2c_2}{3}z^3-\frac{c_2\sqrt{\pi}}{2}z^2\right.\\\nonumber
&\quad\left.+c_2z+2c_2^2z^3e^{z^2}-c_2^2ze^{z^2}+\frac{1}{6}z^2 e^{-z^2}+\frac{5}{12}e^{-z^2}
\right]_{\xi_0}^\xi+\frac{n^2\pi}{4\lambda^2 c_1^2}\int_{\xi_0}^\xi e^{z^2}(2z^2+1)\mathrm{erf}^2(z)\;dz,
\\\nonumber\\
\label{eq:I3}
I_3 = &\frac{2n}{\lambda }\left[  c_2\sqrt{\pi}\mathrm{erfi}(z)+c_2e^{z^2}(2z^2-1)-\frac{1}{4}z^2\left.\right._{2}F_{2}(1,1;-1/2,2;z^2)+\frac{1}{2}z^2\left.\right._{2}F_{2}(1,1;1/2,2;z^2)\right.\\\nonumber
&\quad\left.+\frac{1}{4}z^2 \left.\right._{2}F_{2}(1,1;3/2,2;z^2)-\frac{1}{4} z^4+\frac{1}{3}z^3-\frac{1}{4}z^2+\frac{1}{2}z\right]_{\xi_0}^\xi.
\end{align}
The function $\mathrm{erfi}(z)$ appearing in the components (\ref{eq:I1})-(\ref{eq:I3}) is the imaginary error  function defined by
\be\mathrm{erfi}(z) = -i\cdot \mathrm{erf}(iz),
\ee
and the function $_{p}F_{q}(a_1, a_2, ..., a_p;b_1, b_2, ..., b_q;z)$ is the generalized hypergeometric function \cite{Nikiforov1988} defined by
\be
_{p}F_{q}(a_1, a_2, ..., a_p;b_1, b_2, ..., b_q;z) = \sum_{k=0}^\infty\frac{(a_1)_k (a_2)_k\cdots (a_p)_k}{(b_1)_k (b_2)_k\cdots (b_q)_k}\frac{z^k}{k!},
\ee
where $(a)_k:=a(a+1)\cdots(a+k-1)$ is the Pochhammer symbol.

The components of the surface (\ref{eq:Integrales2}) take the form
\begin{align}\label{eq:ComponentF1}
&F_1 = \frac{1}{2}Re\left[\sqrt{\pi}\left(c_1^2 - \frac{4n^2c_2^2}{\lambda^2c_1^2}\right) \mathrm{erfi}(z)+\left(c_1^2 - \frac{4n^2}{\lambda^2c_1^2} \right)e^{z^2}z(2z^2-1)\right.\\\nonumber
&\quad\quad\left. -\frac{4n^2}{\lambda^2 c_1^2}\left[\frac{\sqrt{\pi}}{2}\left(\frac{1}{3}z^2 \mathrm{erf}(z)+\frac{1}{2}z\cdot \mathrm{erf}(z) +\frac{\sqrt{\pi}}{4} \mathrm{erf}(z)\right)-\frac{c_2}{2}z^2 \left.\right._{2}F_{2}(1,1;-1/2,2;z^2)\right.\right.\\\nonumber
&\quad\quad\left.\left.+c_2\sqrt{\pi}z^2\left.\right._{2}F_{2}(1,1;1/2,2;z^2)+\frac{c_2\sqrt{\pi}}{2}z^2 \left.\right._{2}F_{2}(1,1;3/2,2;z^2)-\frac{c_2\sqrt{\pi}}{2} z^4+\frac{2c_2}{3}z^3-\frac{c_2\sqrt{\pi}}{2}z^2\right.\right.\\\nonumber
&\quad\quad\left.\left.+c_2z+\frac{1}{6}z^2 e^{-z^2}+\frac{5}{12}e^{-z^2}
\right]_{\xi_0}^\xi +\frac{\pi}{16}\int_{\xi_0}^\xi e^{z^2}(2z^2+1)^2 \mathrm{erf}^2(z)dz
\right],
\end{align}
\begin{align}
\label{eq:ComponentF2}
&F_2 = -\frac{1}{2}Im\left[\sqrt{\pi}\left(c_1^2 + \frac{4n^2c_2^2}{\lambda^2c_1^2}\right) \mathrm{erfi}(z)+\left(c_1^2 + \frac{4n^2}{\lambda^2c_1^2} \right)e^{z^2}z(2z^2-1)\right.\\\nonumber
&\quad\quad\left. +\frac{4n^2}{\lambda^2 c_1^2}\left[\frac{\sqrt{\pi}}{2}\left(\frac{1}{3}z^2 \mathrm{erf}(z)+\frac{1}{2}z\cdot \mathrm{erf}(z) +\frac{\sqrt{\pi}}{4} \mathrm{erf}(z)\right)-\frac{c_2}{2}z^2 \left.\right._{2}F_{2}(1,1;-1/2,2;z^2)\right.\right.\\\nonumber
&\quad\quad\left.\left.+c_2\sqrt{\pi}z^2\left.\right._{2}F_{2}(1,1;1/2,2;z^2)+\frac{c_2\sqrt{\pi}}{2}z^2 \left.\right._{2}F_{2}(1,1;3/2,2;z^2)-\frac{c_2\sqrt{\pi}}{2} z^4+\frac{2c_2}{3}z^3-\frac{c_2\sqrt{\pi}}{2}z^2\right.\right.\\\nonumber
&\quad\quad\left.\left.+c_2z+\frac{1}{6}z^2 e^{-z^2}+\frac{5}{12}e^{-z^2}
\right]_{\xi_0}^\xi -\frac{\pi}{16}\int_{\xi_0}^\xi e^{z^2}(2z^2+1)^2 \mathrm{erf}^2(z)dz
\right],
\end{align}
\begin{align}
\label{eq:ComponentF3}
&F_3 = Re\left[\frac{2n}{\lambda}\left[c_2\sqrt{\pi} \mathrm{erfi}(z)+c_2e^{z^2}z(2z^2-1)-\frac{1}{4}z^2\left.\right._{2}F_{2}(1,1;-1/2,2;z^2)\right.\right.\\\nonumber
&\quad\quad\left.\left.+\frac{1}{2}z^2\left.\right._{2}F_{2}(1,1;1/2,2;z^2)+\frac{1}{4}z^2\left.\right._{2}F_{2}(1,1;3/2,2;z^2)-\frac{1}{4}z^4+\frac{1}{3}z^3-\frac{1}{4}z^2+\frac{1}{2}z\right]_{\xi_0}^\xi
\right].
\end{align}
The integral in terms of the error  function (\ref{eq:temp17}) appearing in equations (\ref{eq:ComponentF1}) and (\ref{eq:ComponentF2})
\be\label{eq:integral}
I_4 := \int_{\xi_0}^\xi e^{z^2}(2z^2+1)^2 \mathrm{erf}^2(z)dz
\ee
can be numerically approximated for the purpose of plotting the surface. It may also be reduced to the numerical approximation of the integral
\be
\int_{\xi_0}^\xi e^{z^2} \mathrm{erf}^2(z)dz.
\ee
We integrate (\ref{eq:integral}) by parts by putting $u = \mathrm{erf}^2(z)$ and $dv = e^{z^2}(2z^2+1)^2dz$. The integral $I_4$ becomes
\begin{align}\nonumber
&I_4 = \mathrm{erf}^2(z)\left(\sqrt{\pi}\mathrm{erfi}(z)+e^{z^2}z(2z^2-1)\right)\bigg\vert_{\xi_0}^\xi-\frac{1}{2\sqrt{\pi}}(4z^4-4z^2-1)\mathrm{erf}(z)\bigg\vert_{\xi_0}^\xi\\\label{eq:temp:16}
&\quad\quad-\frac{1}{\pi}e^{-z^2}(2z^3+z)\bigg\vert_{\xi_0}^\xi-4\int_{\xi_0}^\xi e^{-z^2} \mathrm{erf}(z)\mathrm{erfi}(z)dz.
\end{align}
The integral appearing in equation (\ref{eq:temp:16})
\be\label{eq:I5}
I_5:=\int_{\xi_0}^\xi e^{-z^2} \mathrm{erf}(z)\mathrm{erfi}(z)dz
\ee
can also be integrated by parts by putting $s = \mathrm{erfi}(z)$ and $dt = e^{-z^2}\mathrm{erf}(z)dz$. The integral $I_5$ (\ref{eq:I5}) becomes
\be
I_5 = \frac{\sqrt{\pi}}{4}\mathrm{erf}^2(z)\mathrm{erfi}(z)\bigg\vert_{\xi_0}^\xi-\frac{1}{2}\int_{\xi_0}^\xi e^{z^2} \mathrm{erf}^2(z)dz,
\ee
and then the integral $I_4$ (\ref{eq:integral}) becomes
\begin{align}
&I_4 = \left[\mathrm{erf}^2(z)\left(\sqrt{\pi}\mathrm{erfi}(z)+e^{z^2}z(2z^2-1)\right)-\sqrt{\pi}\mathrm{erf}^2(z)\mathrm{erfi}(z)\right.\\\nonumber
&\quad\quad\left.-\frac{1}{2\sqrt{\pi}}(4z^4-4z^2-1)\mathrm{erf}(z)-\frac{1}{\pi}e^{-z^2}(2z^3+z)\right]_{\xi_0}^\xi+2\int_{\xi_0}^\xi e^{z^2} \mathrm{erf}^2(z)dz.
\end{align}
Under the above, we have the following.
\begin{proposition}\label{prop:5}
The formula (\ref{eq:53}) for the immersion of minimal surfaces associated with the general solution $\omega_n(z)$ (\ref{eq:GenSolDecomposed}) of the complex $X_2^{(1)}$-Hermite ODE (\ref{eq:EDOXComplexe}) takes the form
\begin{equation}\label{eq:FTilde_I1I2I3}
 \tilde{F}(n; \lambda; z) = -\frac{i}{2}\left(\begin{array}{cc} I_3+ I_3^* &I_1 - I_2^*  \\ &  \\  -I_2+ I_1^* & -(I_3+ I_3^*) \\ \end{array}\right)\in \frak{su}(2),
 \end{equation}
where $I_k$, $k = 1, 2, 3,$ are the integrals (\ref{eq:I1})-(\ref{eq:I3}).
\end{proposition}
The explicit expressions for the components of the surface $\tilde{F}$ are rather long so we omit them here. The formula (\ref{eq:FTilde_I1I2I3}) is $\frak{su}(2)$-valued because $Tr(\tilde{F}) = 0$ and $\tilde{F}^\dagger = -\tilde{F}$.

Corollary \ref{cor:new} allows us to state the following.
\begin{corollary}
The general solution of the non-degenerate confluent Heun equation (\ref{eq:Heun}) admits a minimal surface representation in $\mathbb{E}^3$.
\end{corollary}
\subsection{Numerical representation of minimal surfaces describing the general \\solution $\omega_n(z)$}

In this section, the 3-dimensional numerical displays of the minimal surfaces (\ref{eq:Integrales2}) are presented in connection with the general solution $\omega_n(z)$ (\ref{eq:GenSolDecomposed}). Even if the $X_2^{(1)}$-Hermite XOPs are not defined for $n=1, 2$, we are able to construct the surfaces describing the behavior of the solutions of the complex $X_2^{(1)}$-Hermite ODE (\ref{eq:EDOXComplexe}) for these values of $n$. This is due to the fact that the surface may be described by the holomorphic wavefunction $\tilde{\Psi}$ (the solution of the linear problem (\ref{eq:52})-(\ref{eq:52_a})), acting as the moving frame on the surface, which is determined by the general solution (\ref{eq:GenSolDecomposed}), defined for all $n \in\mathbb{N}$. For $n=0$, the surface coincides with the plane $F_3\equiv0$. Figures \ref{fig:1} to \ref{fig:3} below show the evolution of the surface for $n=1, 2, 3$. They were obtained using the Mathematica symbolic software and applying the Enneper-Weierstrass immersion formula (\ref{eq:F}) to the general solution $\omega_n(z)$. The components of the surface were calculated in (\ref{eq:ComponentF1}), (\ref{eq:ComponentF2}) and (\ref{eq:ComponentF3}). The integration constants and the parameter were fixed to $c_1=c_2=1$ and $\lambda=\sqrt{\pi}$, respectively. The integration was performed from $\xi_0 = 1+3i$ to $\xi = x+iy$, where $x\in[-1,1]$, $y\in[-1,1]$. The parameter $n$ is related to the complex $X_2^{(1)}$-Hermite ODE (\ref{eq:EDOXComplexe}). As $n$ grows, the surface expands, but the evolution of the surface as the parameter $n$ grows suggests a global flattening phenomena for the third component $F_3$ (notice that there is a change of scale from one figure to another). A mirror symmetry with respect to the plane $F_2\equiv  C$, for some $C<0$, appears clearly in each image.
\begin{figure}[H]
\centering
\includegraphics[width=0.7\textwidth]{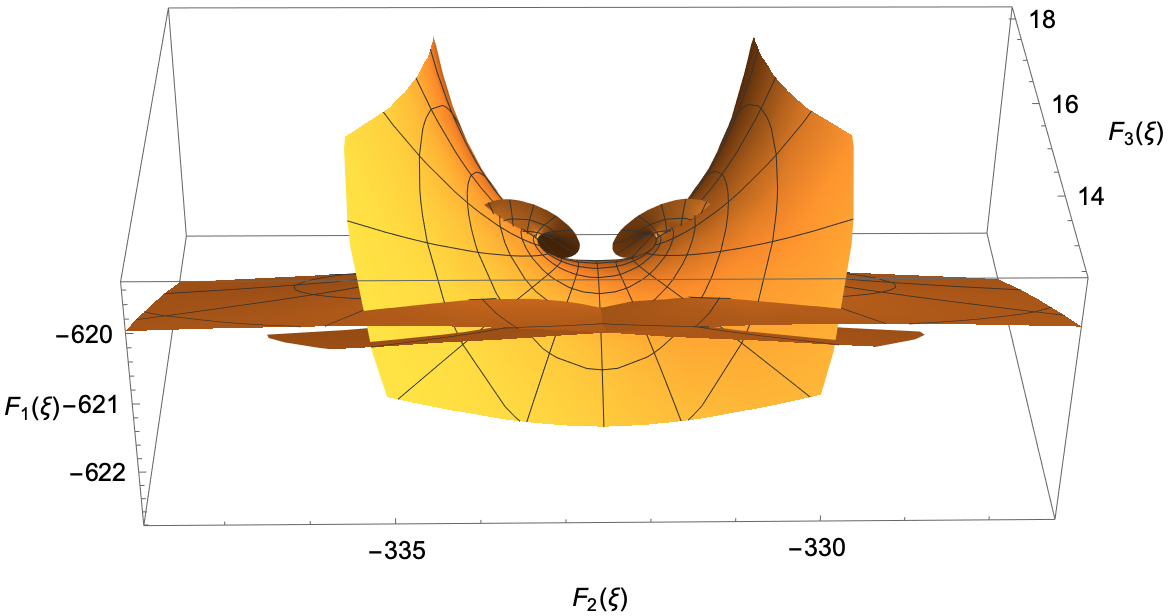}
 \caption{Minimal surface representation of the general solution $\omega_n(z)$, for $n=1$.}
\label{fig:1}
\end{figure}
\begin{figure}[H]
\centering
\includegraphics[width=0.7\textwidth]{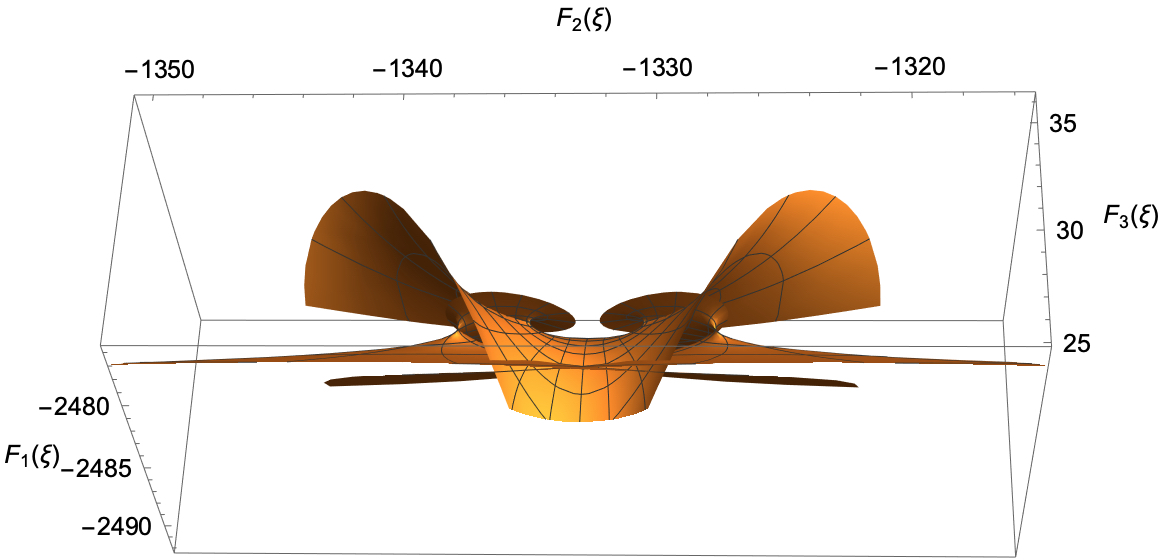}
 \caption{Minimal surface representation of the general solution $\omega_n(z)$, for $n=2$.}
\label{fig:2}
\end{figure}
\begin{figure}[H]
\centering
\includegraphics[width=0.7\textwidth]{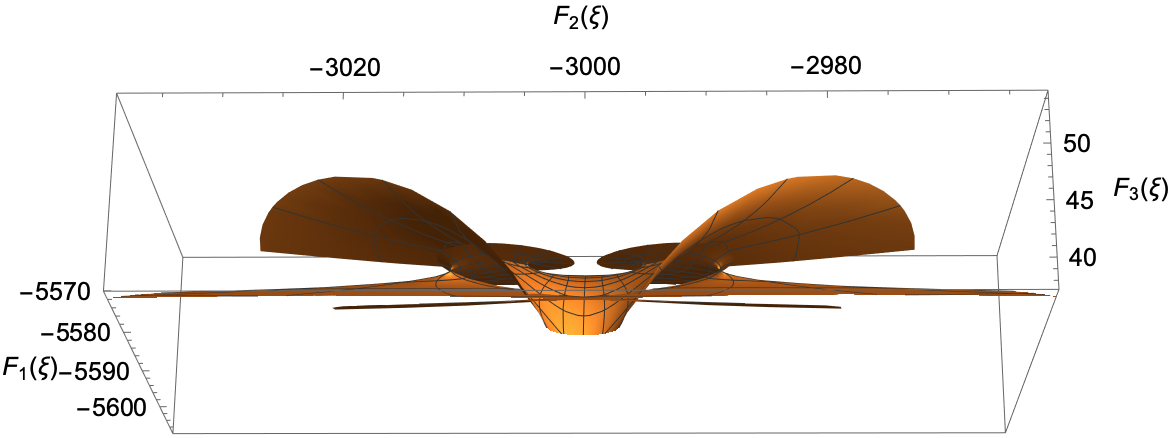}
 \caption{Minimal surface representation of the general solution $\omega_n(z)$, for $n=3$.}
  \label{fig:3}
\end{figure}
\noindent \textbf{Acknowledgements}\\~\\
V.C. and A.M.G. have been partially supported by the Natural Science and Engineering Research Council of Canada (NSERC). The authors thank Jan Derezi{\'n}ski (Department of Mathematical Methods in Physics, Faculty of Physics, University of Warsaw) for helpful comments and discussions on the topic of this paper.
%
\section{Appendix. Proofs}
\subsection{Proof of Proposition \ref{th:GenSol}}\label{app:1}
\small
We now show that the series $\beta_n(z)$ (\ref{eq:Sol1}) is a non-polynomial solution of equation (\ref{eq:EDOXComplexe}) for all $n\in\mathbb{N}$.
\begin{proof}
We proceed by induction. Consider the following proposition:
\be
P_1(n):\;\; \text{The function $\beta_n(z)$ (\ref{eq:Sol1}) is a solution of equation (\ref{eq:EDOXComplexe}) for all }
n\in\mathbb{N}.
\ee
Multiplying by $(1+2z^2)$, equation (\ref{eq:EDOXComplexe}) can be equivalently expressed as
\be\label{eq:EDOXComplexeModif}
\left(1+2z^2\right)\omega''(z)+\left(-4z^3-10z\right)\omega'(z)+\left(2n+4nz^2\right)\omega(z) = 0.
\ee  
Case $n = 0$. Equation (\ref{eq:EDOXComplexeModif}) becomes
\be\label{eq:EDOXComplexeModifZero}
\left(1+2z^2\right)\omega''(z)+\left(-4z^3-10z\right)\omega'(z) = 0.
\ee 
Let us denote
\be\label{eq:Delta1}
\Delta_1(k):=(-1)^k2^{k-1}((2(k-1))^2+1)\prod_{j=1}^{k-2}(1-2(1+j)).
\ee
Then the series (\ref{eq:Sol1}) and its first and second-order derivatives take the form
\begin{align}\label{eq:H0}
\beta_0(z) &= 1+z+\frac{5}{3}z^3+\sum_{k=3}^\infty\frac{\Delta_1(k)}{(2k-1)!}z^{2k-1},\\\label{eq:H0Derivee}
\frac{d\beta_0}{dz}(z) &= 1+5z^2+\sum_{k=3}^\infty\frac{\Delta_1(k)}{(2k-2)!}z^{2k-2},\qquad
\frac{d^2\beta_0}{dz^2}(z) = 10z+\sum_{k=3}^\infty\frac{\Delta_1(k)}{(2k-3)!}z^{2k-3}.
\end{align}
Substituting (\ref{eq:H0Derivee}) into the left-hand side (LHS) of equation (\ref{eq:EDOXComplexeModifZero}), we obtain
\begin{align}\nonumber
&G_1(0;z) = \left(1+2z^2\right)\cdot10z+\left(-4z^3-10z\right)\cdot\left(1+5z^2\right)+\sum_{k=3}^\infty\frac{\Delta_1(k)}{(2k-3)!}z^{2k-3}\\\label{eq:tempnew}
&\quad+\sum_{k=3}^\infty\frac{2\Delta_1(k)}{(2k-3)!}z^{2k-1}-\sum_{k=3}^\infty\frac{4\Delta_1(k)}{(2k-2)!}z^{2k+1}-\sum_{k=3}^\infty\frac{10\Delta_1(k)}{(2k-2)!}z^{2k-1}.
\end{align}
In order to obtain powers corresponding to $(2k-1)$ in all series, we perform a translation of the summation variable where necessary. Extracting the terms of degree $k\leq3$, and considering the terms outside of a series in equation (\ref{eq:tempnew}), we see that they cancel each other. Regrouping all series, we get
\begin{align}\label{eq:G1}
G_1(0;z)=&\sum_{k=4}^\infty\left[\frac{\Delta_1(k+1)}{(2k-1)!}
+\frac{2\Delta_1(k)}{(2k-3)!}
-\frac{4\Delta_1(k-1)}{(2k-4)!}  
-\frac{10\Delta_1(k)}{(2k-2)!} \right]z^{2k-1}.
\end{align}
Evaluating $\Delta_1$ from relation (\ref{eq:Delta1}), we obtain
\begin{align}
G_1(0;z) = \sum_{k=4}^\infty\left[   \frac{(-1)^k2^{k}\prod_{j=1}^{k-3}(1+2(1+j))}{(2k-4)!}\cdot\Delta_2(k) \right]z^{2k-1},
\end{align}
where
\begin{align}\label{eq:Delta2}
&\Delta_2(k):= - \frac{((2k)^2+1)(1-2(1+(k-2)))(1-2(1+(k-1)))}{(2k-3)(2k-2)(2k-1)}+((2(k-2))^2+1)\\\nonumber
&\qquad\qquad
+\frac{((2(k-1))^2+1)(1-2(1+(k-2)))}{(2k-3)} -\frac{5((2(k-1))^2+1)(1-2(1+(k-2)))}{(2k-3)(2k-2)}
=0,
\end{align}
for all $k\geq4$. We conclude that $P_1(0)$ is true.

Applying the induction hypothesis, we suppose that $P_1(n-1)$ is true for some $n\geq1$, \textit{i.e.}
\begin{align}\label{eq:EDOXComplexeModifNMoins1}
&\left(1+2z^2\right)\left(\beta_{n-1}(z)\right)''+\left(-4z^3-10z\right)\left(\beta_{n-1}(z)\right)'+2(n-1)\left(1+2z^2\right)\beta_{n-1}(z) =0.
\end{align}
We want to show that $P_1(n)$ is also true, \textit{i.e.}
\begin{align}\label{eq:EDOXComplexeModifN}
&\left(1+2z^2\right)\left(\beta_{n}(z)\right)''+\left(-4z^3-10z\right)\left(\beta_{n}(z)\right)'+2n\left(1+2z^2\right)\beta_{n}(z) =0.
\end{align}
Subtracting the LHS of (\ref{eq:EDOXComplexeModifN}) from (\ref{eq:EDOXComplexeModifNMoins1}), we obtain the equality
\begin{align}\label{eq:EDOAMONTRER}
& \left(1+2z^2\right)\left[\left(\beta_{n-1}\right)''-\left(\beta_{n}\right)''\right](z)+\left(-4z^3-10z\right)\left[\left(\beta_{n-1}\right)'-\left(\beta_{n}\right)'\right](z)\\\nonumber
&\qquad\qquad\qquad\qquad\qquad\qquad\qquad+2n(1+2z^2)\left[\beta_{n-1}-\beta_{n}\right](z) -2(1+2z^2)\beta_{n-1}(z)=0
\end{align}
and need to prove that it is true. Let us define
\begin{align}\label{eq:Delta7}
\Delta_3(k):&=(n-1)(n-((2k-1)^2+2))\prod_{j=1}^{k-2}(n-(2(1+j)+1)),\\\label{eq:Delta8}
\Delta_4(k):&=n(n-((2k-1)^2+1))\prod_{j=1}^{k-2}(n-2(1+j)),\\\label{eq:Delta9}
\Delta_{5}(k):&=(n-((2(k-1))^2+2))\prod_{j=1}^{k-2}(n-2(1+j)),\\\label{eq:Delta10}
\Delta_{6}(k):&=(n-((2(k-1))^2+1))\prod_{j=1}^{k-2}(n-2(1+j)+1).
\end{align}
Then we get
\begin{align}\label{eq:Diff1}
&\left[\beta_{n-1}-\beta_{n}\right](z)  = z^2+\frac{1}{3}z^3+\frac{11-2n}{6}z^4+\sum_{k=3}^\infty\left[\frac{(-1)^k2^k}{(2k)!}\left(\Delta_3 - \Delta_4\right)z^{2k}+\frac{(-1)^{k+1}2^{k-1}}{(2k-1)!}\left(\Delta_5 - \Delta_{6}\right)z^{2k-1}\right],\\\label{eq:Diff2}
&\left[\left(\beta_{n-1}\right)'-\left(\beta_{n}\right)'\right](z)=2z+z^2+\frac{22-4n}{3}z^3+\sum_{k=3}^\infty\left[  \frac{(-1)^k2^k}{(2k-1)!}\left(\Delta_3 - \Delta_4\right)z^{2k-1} + \frac{(-1)^{k+1}2^{k-1}}{(2k-2)!}\left(\Delta_5 - \Delta_{6}\right)z^{2k-2} \right],\\\label{eq:Diff3}
&\left[\left(\beta_{n-1}\right)''-\left(\beta_{n}\right)''\right](z)=2+2z+(22-4n)z^2+\sum_{k=3}^\infty\left[\frac{(-1)^k2^k}{(2k-2)!}\left(\Delta_3 - \Delta_4\right)z^{2k-2}+\frac{(-1)^{k+1}2^{k-1}}{(2k-3)!}\left(\Delta_5 - \Delta_{6}\right)z^{2k-3}\right].
\end{align}
Substituting (\ref{eq:Diff1})-(\ref{eq:Diff3}) into the LHS of equation (\ref{eq:EDOAMONTRER}), we obtain
\begin{align}\nonumber
&G_1(n;z) =\left(1+2z^2\right)\left(2+2z+(22-4n)z^2\right)+\left(-4z^3-10z\right)   \left(2z+z^2+\frac{22-4n}{3}z^3\right)\\\nonumber
&\qquad+2n(1+2z^2)\left(z^2+\frac{1}{3}z^3+\frac{11-2n}{6}z^4\right)-2\left(1+2z^2\right)\left(1+z-(n-1)z^2-\frac{n-6}{3}z^3+\frac{(n-1)(n-11)}{6}z^4\right)\\\nonumber
&\qquad+\sum_{k=3}^\infty\left[  \frac{(-1)^{k}2^k}{(2k-2)!}\left(\Delta_3 - \Delta_4\right)z^{2k-2} +\frac{(-1)^{k+1}2^{k-1}}{(2k-3)!}\left(\Delta_5 - \Delta_6\right)z^{2k-3}\right]\\\nonumber
&\qquad+\sum_{k=3}^\infty\left[  \frac{(-1)^{k}2^{k+1}}{(2k-2)!}\left(\Delta_3 - \Delta_4\right)z^{2k} +\frac{(-1)^{k+1}2^{k}}{(2k-3)!}\left(\Delta_5 - \Delta_6\right)z^{2k-1}\right]\\\nonumber
&\qquad+\sum_{k=3}^\infty\left[  \frac{(-1)^{k+1}2^{k+2}}{(2k-1)!}\left(\Delta_3 - \Delta_4\right)z^{2k} +\frac{(-1)^{k}2^{k+1}}{(2k-2)!}\left(\Delta_5 - \Delta_6\right)z^{2k+1}\right]\\
\label{eq:TempP4}
&\qquad+\sum_{k=3}^\infty\left[  \frac{5(-1)^{k+1}2^{k+1}}{(2k-1)!}\left(\Delta_3 - \Delta_4\right)z^{2k} +\frac{5(-1)^{k}2^{k}}{(2k-2)!}\left(\Delta_5 - \Delta_6\right)z^{2k-1}\right]\\\nonumber
&\qquad+\sum_{k=3}^\infty\left[  \frac{(-1)^{k}2^{k+1}n}{(2k)!}\left(\Delta_3 - \Delta_4\right)z^{2k} +\frac{(-1)^{k+1}2^{k}n}{(2k-1)!}\left(\Delta_5 - \Delta_6\right)z^{2k-1}\right]
\\\nonumber
&\qquad+\sum_{k=3}^\infty\left[  \frac{(-1)^{k}2^{k+2}n}{(2k)!}\left(\Delta_3 - \Delta_4\right)z^{2k+2} +\frac{(-1)^{k+1}2^{k+1}n}{(2k-1)!}\left(\Delta_5 - \Delta_6\right)z^{2k+1}\right]\\\nonumber
&\qquad+\sum_{k=3}^\infty\left[  \frac{(-1)^{k+1}2^{k+1}}{(2k)!}\Delta_{3}z^{2k} +\frac{(-1)^{k}2^{k}}{(2k-1)!}\Delta_{5}z^{2k-1}\right]+\sum_{k=3}^\infty\left[  \frac{(-1)^{k+1}2^{k+2}}{(2k)!}\Delta_{3}z^{2k+2} +\frac{(-1)^{k}2^{k+1}}{(2k-1)!}\Delta_{5}z^{2k+1}\right].
\end{align}
In order to obtain powers corresponding to $(2k)$ and $(2k-1)$ in all series, we perform a translation of the summation variable where necessary. Extracting the terms of degree $k\leq3$, and considering the terms outside of a series in equation (\ref{eq:TempP4}), we see that they cancel each other.  Regrouping all series, we obtain
\begin{align}\nonumber
&G_1(n;z) =\sum_{k=4}^\infty\left[  \frac{(-1)^{k}2^{k+1}}{(2k-3)!}\left(\prod_{j=1}^{k-3}\left( n-(2(1+j)+1)\right)\Delta_{7}-\prod_{j=1}^{k-3}\left( n-2(1+j)\right)\Delta_{8}\right)z^{2k}\right. \\
& \qquad\qquad+\left.\frac{(-1)^{k+1}2^{k}}{(2k-4)!}\left(\prod_{j=1}^{k-3}\left( n-2(1+j)\right)\Delta_{9}-\prod_{j=1}^{k-3}\left( n-(2(1+j)+1)\right)\Delta_{10}\right) z^{2k-1} \right],
\end{align}
where
\begin{align}\label{eq:Delta7Prime}
\Delta_{7}(k):&=-\frac{(n-1)(n-((2k+1)^2+2))(n-(2(k-1)+1))(n-(2k+1))}{(2k-2)(2k-1)(2k)}\\\nonumber
&+(n-1)(n-((2k-1)^2+2))(n-(2(k-1))+1)\cdot\frac{(2k-1)(2k)-5(2k)+n}{(2k-2)(2k-1)(2k)}\\\nonumber
&-\frac{(n-1)(n-((2k-1)^2+2))(n-(2(k-1)+1))}{(2k-2)(2k-1)(2k)}+\frac{(n-1)(n-((2k-3)^2+2))}{(2k-2)}\\\nonumber
&+(n-1)(n-((2k-3)^2+2))\left(1-\frac{n}{(2k-2)}\right)=0,
\end{align}
\begin{align}
\label{eq:Delta8Prime}
\Delta_{8}(k):&=-\frac{n(n-((2k+1)^2+1))(n-2(k-1))(n-2k)}{(2k-2)(2k-1)(2k)}\\\nonumber
&+n(n-((2k-1)^2+1))(n-2(k-1))\cdot\frac{(2k-1)(2k)-5(2k)+n}{(2k-2)(2k-1)(2k)}
\\\nonumber
&+n(n-((2k-3)^2+1))\left(1-\frac{n}{(2k-2)}\right)=0,\\
\label{eq:Delta13}
\Delta_{9}(k):&=-\frac{(n-((2k)^2+2))(n-2(k-1))(n-2k)}{(2k-3)(2k-2)(2k-1)}+\frac{(n-((2(k-2))^2+2))}{(2k-3)}\\\nonumber
&+(n-((2(k-1))^2+2))(n-2(k-1))\cdot \frac{(2k-2)(2k-1)-5(2k-1)+n}{(2k-3)(2k-2)(2k-1)}\\\nonumber
&+(n-((2(k-2))^2+2))\cdot\frac{2k-3-n}{(2k-3)}-\frac{(n-((2(k-1))^2+2))(n-2(k-1))}{(2k-3)(2k-2)(2k-1)} = 0,\\
\label{eq:Delta14}
\Delta_{10}(k):&=-\frac{(n-((2k)^2+1))(n-2(k-1)+1)(n-2k+1)}{(2k-3)(2k-2)(2k-1)}\\\nonumber
&+(n-((2(k-1))^2+1))(n-2(k-1)+1)\cdot \frac{(2k-2)(2k-1)-5(2k-1)+n}{(2k-3)(2k-2)(2k-1)}\\\nonumber
&+(n-((2(k-2))^2+1))\left(1-\frac{n}{(2k-3)}\right)=0,
\end{align}
for all $k\geq4$. Thus we conclude that $P_1(n)$ is true.

By construction, the solution $\beta_n(z)$ (\ref{eq:Sol1}) is non-polynomial, because the coefficients $c_{2k}(n)$ (\ref{eq:coeffPair}) and $c_{2k-1}(n)$ (\ref{eq:coeffImpair}), associated with even and odd powers of $z$, respectively, are polynomials of the parameter $n$, possessing no root on the gap sequence (see Table \ref{tab:2}). This completes the proof of Proposition \ref{th:GenSol}.
\end{proof}
\subsection{Proof of Proposition \ref{th:GenSol1}}\label{app:2}
We now show that the series $\mu_{n}(z)$ (\ref{eq:mu}) is a polynomial solution of equation (\ref{eq:EDOXComplexe}) for all $n\in2\mathbb{N}\backslash\{2\}$ and that the series $\nu_{n}(z)$ (\ref{eq:Sol2}) is a polynomial solution of equation (\ref{eq:EDOXComplexe}) for all $n\in(2\mathbb{N}-1)\backslash\{1\}$.
\begin{proof}
 We proceed by induction. Consider the following proposition:
\be
P_2(n):\;\; \text{The function $\mu_n(z)$ (\ref{eq:mu}) is a solution of equation (\ref{eq:EDOXComplexe}) for all }
n\in\mathbb{N}.
\ee 
Case $n = 0$. 
The series $\mu_n(z)$ (\ref{eq:mu}) and its first and second-order derivatives take the form
\begin{align}
\mu_0(z) = 1, \qquad
\frac{d\mu_0}{dz}(z) = 0,\qquad
\frac{d^2\mu_0}{dz^2}(z) = 0,
\end{align}
so we see immediately that $\mu_0(z)$ is a solution of equation (\ref{eq:EDOXComplexeModifZero}). We conclude that $P_2(0)$ is true.

Applying the induction hypothesis, we suppose that $P_2(n-1)$ is also true for some $n\geq1$, \textit{i.e.}
\begin{align}\label{eq:EDOXComplexeModifNMoins1mu}
&\left(1+2z^2\right)\left(\mu_{n-1}(z)\right)''+\left(-4z^3-10z\right)\left(\mu_{n-1}(z)\right)'+2(n-1)\left(1+2z^2\right)\mu_{n-1}(z) =0.
\end{align}
We want to show that $P_2(n)$ is true, \textit{i.e.}
\begin{align}\label{eq:EDOXComplexeModifNmu}
&\left(1+2z^2\right)\left(\mu_{n}(z)\right)''+\left(-4z^3-10z\right)\left(\mu_{n}(z)\right)'+2n\left(1+2z^2\right)\mu_{n}(z) =0.
\end{align}
Subtracting the LHS of (\ref{eq:EDOXComplexeModifNmu}) from (\ref{eq:EDOXComplexeModifNMoins1mu}), we obtain the equality
\begin{align}\label{eq:EDOAMONTRERmu}
& \left(1+2z^2\right)\left[\left(\mu_{n-1}\right)''-\left(\mu_{n}\right)''\right](z)+\left(-4z^3-10z\right)\left[\left(\mu_{n-1}\right)'-\left(\mu_{n}\right)'\right](z)\\\nonumber
&\qquad\qquad\qquad\qquad\qquad\qquad\qquad+2n(1+2z^2)\left[\mu_{n-1}-\mu_{n}\right] -2(1+2z^2)\mu_{n-1}(z)=0
\end{align}
and need to show that it is true. We get
\begin{align}\label{eq:Diff1mu}
&\left[\mu_{n-1}-\mu_{n}\right](z)  = z^2+\frac{11-2n}{6}z^4+\sum_{k=3}^\infty\left[\frac{(-1)^k2^k}{(2k)!}\left(\Delta_3 - \Delta_4\right)z^{2k}+\right],\\\label{eq:Diff2mu}
&\left[\left(\mu_{n-1}\right)'-\left(\mu_{n}\right)'\right](z)=2z+\frac{22-4n}{3}z^3+\sum_{k=3}^\infty\left[  \frac{(-1)^k2^k}{(2k-1)!}\left(\Delta_3 - \Delta_4\right)z^{2k-1} \right],\\\label{eq:Diff3mu}
&\left[\left(\mu_{n-1}\right)''-\left(\mu_{n}\right)''\right](z)=2+(22-4n)z^2+\sum_{k=3}^\infty\left[\frac{(-1)^k2^k}{(2k-2)!}\left(\Delta_3 - \Delta_4\right)z^{2k-2}\right].
\end{align}
Substituting (\ref{eq:Diff1mu})-(\ref{eq:Diff3mu}) into the LHS of equation (\ref{eq:EDOAMONTRERmu}), we obtain
\begin{align}\nonumber
&G_2(n;z) =\left(1+2z^2\right)\left(2+(22-4n)z^2\right)+\left(-4z^3-10z\right)   \left(2z+\frac{22-4n}{3}z^3\right)\\\label{eq:TempP4mu}
&\qquad+2n(1+2z^2)\left(z^2+\frac{11-2n}{6}z^4\right)-2\left(1+2z^2\right)\left(1-(n-1)z^2+\frac{(n-1)(n-11)}{6}z^4\right)\\\nonumber
&\qquad+\sum_{k=3}^\infty\left[  \frac{(-1)^{k}2^k}{(2k-2)!}\left(\Delta_3 - \Delta_4\right)z^{2k-2} \right]+\sum_{k=3}^\infty\left[  \frac{(-1)^{k}2^{k+1}}{(2k-2)!}\left(\Delta_3 - \Delta_4\right)z^{2k} \right]+\sum_{k=3}^\infty\left[  \frac{(-1)^{k+1}2^{k+2}}{(2k-1)!}\left(\Delta_3 - \Delta_4\right)z^{2k} \right]\\\nonumber
&\qquad+\sum_{k=3}^\infty\left[  \frac{5(-1)^{k+1}2^{k+1}}{(2k-1)!}\left(\Delta_3 - \Delta_4\right)z^{2k} \right]+\sum_{k=3}^\infty\left[  \frac{(-1)^{k}2^{k+1}n}{(2k)!}\left(\Delta_3 - \Delta_4\right)z^{2k} \right]+\sum_{k=3}^\infty\left[  \frac{(-1)^{k}2^{k+2}n}{(2k)!}\left(\Delta_3 - \Delta_4\right)z^{2k+2} \right]\\\nonumber
&\qquad+\sum_{k=3}^\infty\left[  \frac{(-1)^{k+1}2^{k+1}}{(2k)!}\Delta_{3}z^{2k} \right]+\sum_{k=3}^\infty\left[  \frac{(-1)^{k+1}2^{k+2}}{(2k)!}\Delta_{3}z^{2k+2} \right].
\end{align}
In order to obtain powers corresponding to $(2k)$ in all series, we perform a translation of the summation variable where necessary. Extracting the terms of degree $k\leq3$, and considering the terms outside of a series in equation (\ref{eq:TempP4mu}), we see that they cancel each other.  Regrouping all series, we obtain
\begin{align}\nonumber
&G_2(n;z) =\sum_{k=4}^\infty\left[  \frac{(-1)^{k}2^{k+1}}{(2k-3)!}\left(\prod_{j=1}^{k-3}\left( n-(2(1+j)+1)\right)\Delta_{7}-\prod_{j=1}^{k-3}\left( n-2(1+j)\right)\Delta_{8}\right)z^{2k} \right],
\end{align}
where we already showed by the relations (\ref{eq:Delta7Prime}) and (\ref{eq:Delta8Prime}) that $\Delta_7(k)\equiv\Delta_8(k)\equiv0$ for all $k\geq4$. Thus we conclude that $P_2(n)$ is true.

By construction, the coefficients $c_{2k}(n)$ of the series $\mu_n(z)$ (\ref{eq:mu}) possess only even roots $\lambda_p(k)$  from which $n=2$ is excluded, as illustrated in Table \ref{tab:2} and by the polynomials $p_k(n)$ (\ref{eq:pk}). Therefore the only polynomial cases are $\mu_{2l}$, where $l\in\{0,2, 3, 4...\}$.
\\~\\
Consider the following proposition:
\be
P_3(n):\;\; \text{The function $\nu_n(z)$ (\ref{eq:Sol2}) is a solution of equation (\ref{eq:EDOXComplexe}) for all }
n\in\mathbb{N}.
\ee
Case $n = 0$. The series $\nu_n(z)$ (\ref{eq:Sol2}) and its first and second-order derivatives take the form
\begin{align}\label{eq:H0Prime}
\nu_0(z) &= z+\frac{5}{3}z^3+\sum_{k=3}^\infty\frac{\Delta_1(k)}{(2k-1)!}z^{2k-1},\\\label{eq:H0DeriveePrime}
\frac{d\nu_0}{dz}(z) &= 1+5z^2+\sum_{k=3}^\infty\frac{\Delta_1(k)}{(2k-2)!}z^{2k-2},\qquad
\frac{d^2\nu_0}{dz^2}(z) = 10z+\sum_{k=3}^\infty\frac{\Delta_1(k)}{(2k-3)!}z^{2k-3}.
\end{align}
Substituting (\ref{eq:H0DeriveePrime}) into equation (\ref{eq:EDOXComplexeModifZero}), we obtain
\begin{align}\nonumber
&G_3(0;z)=\left(1+2z^2\right)\cdot10z+\left(-4z^3-10z\right)\cdot\left(1+5z^2\right)+\sum_{k=3}^\infty\frac{\Delta_1(k)}{(2k-3)!}z^{2k-3}\\\label{eq:tempnewnew}
&\qquad\qquad+\sum_{k=3}^\infty\frac{2\Delta_1(k)}{(2k-3)!}z^{2k-1}-\sum_{k=3}^\infty\frac{4\Delta_1(k)}{(2k-2)!}z^{2k+1}-\sum_{k=3}^\infty\frac{10\Delta_1(k)}{(2k-2)!}z^{2k-1}.
\end{align}
In order to obtain powers corresponding to $(2k-1)$ in all series, we perform a translation of the summation variable where necessary. Extracting the terms of degree $k\leq3$, and considering the terms outside of a series in equation (\ref{eq:tempnewnew}), we see that they cancel each other. Regrouping all series, we get
\begin{align}
&G_3(0;z)=\sum_{k=4}^\infty\left[\frac{\Delta_1(k+1)}{(2k-1)!}
+\frac{2\Delta_1(k)}{(2k-3)!}
-\frac{4\Delta_1(k-1)}{(2k-4)!}  
-\frac{10\Delta_1(k)}{(2k-2)!} \right]z^{2k-1}.
\end{align}
Evaluating $\Delta_1$ from relation (\ref{eq:Delta1}), we obtain
\begin{align}
G_3(0;z)=\sum_{k=4}^\infty\left[   \frac{(-1)^k2^{k}\prod_{j=1}^{k-3}(1+2(1+j))}{(2k-4)!}\cdot\Delta_2(k) \right]z^{2k-1}.
\end{align}
We already showed by the relation (\ref{eq:Delta2}) that $\Delta_2(k)\equiv0$ for all $k\geq4$. We conclude that $P_3(0)$ is true as well.

Applying the induction hypothesis, we suppose that $P_3(n-1)$ is true for some $n\geq1$, \textit{i.e.}
\begin{align}\label{eq:EDOXComplexeModifNMoins1Prime}
&\left(1+2z^2\right)\left(\nu_{n-1}(z)\right)''+\left(-4z^3-10z\right)\left(\nu_{n-1}(z)\right)'+2(n-1)\left(1+2z^2\right)\nu_{n-1}(z) =0.
\end{align}
We want to show that $P_3(n)$ is true, \textit{i.e.}
\begin{align}\label{eq:EDOXComplexeModifNPrime}
&\left(1+2z^2\right)\left(\nu_{n}(z)\right)''+\left(-4z^3-10z\right)\left(\nu_{n}(z)\right)'+2n\left(1+2z^2\right)\nu_{n}(z) =0.
\end{align}
Subtracting the left-hand side (LHS) of (\ref{eq:EDOXComplexeModifNPrime}) from (\ref{eq:EDOXComplexeModifNMoins1Prime}), we obtain the equality
\begin{align}\label{eq:EDOAMONTRERPrime}
&\left(1+2z^2\right)\left[\left(\nu_{n-1}\right)''-\left(\nu_{n}\right)''\right](z)+\left(-4z^3-10z\right)\left[\left(\nu_{n-1}\right)'-\left(\nu_{n}\right)'\right](z)\\\nonumber
&\qquad\qquad\qquad\qquad\qquad+2n\left(1+2z^2\right)\left[\nu_{n-1}-\nu_{n}\right](z) -2(1+2z^2)\nu_{n-1}(z)=0
\end{align}
and need to prove that it is true. We get
\begin{align}\label{eq:Diff1Prime}
&\left[\nu_{n-1}-\nu_{n}\right](z)  =\frac{1}{3}z^3+\sum_{k=3}^\infty\left[\frac{(-1)^{k+1}2^{k-1}}{(2k-1)!}\left(\Delta_5 - \Delta_{6}\right)z^{2k-1}\right],\\\label{eq:Diff2Prime}
&\left[\left(\nu_{n-1}\right)'-\left(\nu_{n}\right)'\right](z)=z^2+\sum_{k=3}^\infty\left[  \frac{(-1)^{k+1}2^{k-1}}{(2k-2)!}\left(\Delta_5 - \Delta_{6}\right)z^{2k-2} \right],\\\label{eq:Diff3Prime}
&\left[\left(\nu_{n-1}\right)''-\left(\nu_{n}\right)''\right](z)=2z+\sum_{k=3}^\infty\left[\frac{(-1)^{k+1}2^{k-1}}{(2k-3)!}\left(\Delta_5 - \Delta_{6}\right)z^{2k-3}\right].
\end{align}
Substituting (\ref{eq:Diff1Prime})-(\ref{eq:Diff3Prime}) into the LHS of equation (\ref{eq:EDOAMONTRERPrime}), we obtain
\begin{align}\label{eq:TempP4Prime}
&G_3(n;z) = \left(1+2z^2\right)\cdot2z+\left(-4z^3-10z\right)   \cdot z^2+2n(1+2z^2)\cdot \frac{1}{3}z^3-2\left(1+2z^2\right)\left(z-\frac{n-6}{3}z^3\right)\\\nonumber
&\qquad+\sum_{k=3}^\infty\left[  \frac{(-1)^{k+1}2^{k-1}}{(2k-3)!}\left(\Delta_5 - \Delta_{6}\right)z^{2k-3}\right]+\sum_{k=3}^\infty\left[  \frac{(-1)^{k+1}2^{k}}{(2k-3)!}\left(\Delta_5 - \Delta_{6}\right)z^{2k-1}\right]+\sum_{k=3}^\infty\left[ \frac{(-1)^{k}2^{k+1}}{(2k-2)!}\left(\Delta_5 - \Delta_{6}\right)z^{2k+1}\right]\\\nonumber
&\qquad+\sum_{k=3}^\infty\left[ \frac{5(-1)^{k}2^{k}}{(2k-2)!}\left(\Delta_5 - \Delta_{6}\right)z^{2k-1}\right]+\sum_{k=3}^\infty\left[  \frac{(-1)^{k+1}2^{k}n}{(2k-1)!}\left(\Delta_5 - \Delta_{6}\right)z^{2k-1}\right]+\sum_{k=3}^\infty\left[ \frac{(-1)^{k+1}2^{k+1}n}{(2k-1)!}\left(\Delta_5 - \Delta_{6}\right)z^{2k+1}\right]\\\nonumber
&\qquad+\sum_{k=3}^\infty\left[ \frac{(-1)^{k}2^{k}}{(2k-1)!}\Delta_{5}z^{2k-1}\right]+\sum_{k=3}^\infty\left[  \frac{(-1)^{k}2^{k+1}}{(2k-1)!}\Delta_{5}z^{2k+1}\right].
\end{align}
In order to obtain powers corresponding to $(2k-1)$ in all series, we perform a translation of the summation variable where necessary. Extracting the terms of degree $k\leq3$, and considering the terms outside of a series in (\ref{eq:TempP4Prime}), we see that they cancel each other.  Regrouping all series, we obtain
\begin{align}
&G_3(n;z) =\sum_{k=4}^\infty\left[  \frac{(-1)^{k+1}2^{k}}{(2k-4)!}\left(\prod_{j=1}^{k-3}\left( n-2(1+j)\right)\Delta_{9}-\prod_{j=1}^{k-3}\left( n-(2(1+j)+1)\right)\Delta_{10}\right) z^{2k-1} \right]
\end{align}
where we already showed by the relations (\ref{eq:Delta13}) and (\ref{eq:Delta14}) that $\Delta_{9}(k)\equiv\Delta_{10}(k)\equiv0$ for all $k\geq4$. We conclude that $P_3(n)$ is true.

By construction, the coefficients $\tilde{c}_k(n)$ of the series $\nu_n(z)$ (\ref{eq:Sol2}) possess only odd roots $\lambda_q(k)$ from which $n=1$ is excluded, as illustrated by Table \ref{tab:2} and by the polynomials $q_k(n)$ (\ref{eq:qk}). Therefore the only polynomial cases are $\nu_{2l-1}$, where $l\geq2$, which completes the proof of Proposition \ref{th:GenSol1}.
\end{proof}
\bibliographystyle{spmpsci}      

\begin{thebibliography}{10}
\providecommand{\url}[1]{{#1}}
\providecommand{\urlprefix}{URL }
\expandafter\ifx\csname urlstyle\endcsname\relax
  \providecommand{\doi}[1]{DOI~\discretionary{}{}{}#1}\else
  \providecommand{\doi}{DOI~\discretionary{}{}{}\begingroup
  \urlstyle{rm}\Url}\fi

\bibitem{Adler1994}
Adler, V.{\'{E}}.: {A modification of Crum's method}.
\newblock Theoretical and Mathematical Physics \textbf{101}(3), 1381--1386
  (1994)

\bibitem{Bobenko1994}
Bobenko, A.I.: {Surfaces in terms of 2 by 2 matrices. Old and new integrable
  cases}.
\newblock Harmonic maps and integrable systems pp. 83--127 (1994).
\newblock \doi{10.1007/978-3-663-14092-4-5}

\bibitem{Bobenko2000}
Bobenko, A.I., Eitner, U.: {Painlev{\'{e}} Equations in the Differential
  Geometry of Surfaces}.
\newblock Springer-Verlag, Berlin (2000)

\bibitem{Bonneux2019}
Bonneux, N.: {Exceptional Jacobi polynomials}.
\newblock Journal of Approximation Theory \textbf{239}, 72--112 (2019).
\newblock \doi{10.1016/j.jat.2018.11.002}

\bibitem{Bonneux2018a}
Bonneux, N., Kuijlaars, A.B.: {Exceptional Laguerre Polynomials}.
\newblock Studies in Applied Mathematics \textbf{141}(4), 547--595 (2018).
\newblock \doi{10.1111/sapm.12204}

\bibitem{Bonneux2018}
Bonneux, N., Stevens, M.: {Recurrence relations for Wronskian Hermite
  polynomials}.
\newblock Symmetry, Integrability and Geometry: Methods and Applications
  (SIGMA) \textbf{14}(048) (2018).
\newblock \doi{10.3842/SIGMA.2018.048}

\bibitem{Cariena2008}
Cari{\~{n}}ena, J.F., Perelomov, A.M., Ra{\~{n}}ada, M.F., Santander, M.: {A
  quantum exactly solvable nonlinear oscillator related to the isotonic
  oscillator}.
\newblock Journal of Physics A: Mathematical and Theoretical \textbf{41}(8)
  (2008).
\newblock \doi{10.1088/1751-8113/41/8/085301}

\bibitem{Chalifour2019}
Chalifour, V., Grundland, A.M.: {Minimal surfaces associated with orthogonal
  polynomials (accepted for publication, 2019)}.
\newblock Journal of Nonlinear Mathematical Physics, arXiv:1912.10899v1

\bibitem{Derezinski2020}
Derezi{\'{n}}ski, J., Latosi{\'{n}}ski, A., Ishkhanyan, A.: {From Heun class
  equations to Painlev{\'{e}} equations}.
\newblock arXiv:2007.05698  (2020)

\bibitem{Dimitrov2014}
Dimitrov, D.K., Lun, Y.C.: {Monotonicity, interlacing and electrostatic
  interpretation of zeros of exceptional Jacobi polynomials}.
\newblock Journal of Approximation Theory \textbf{181}, 18--29 (2014).
\newblock \doi{10.1016/j.jat.2014.01.007}

\bibitem{Doliwa2012}
Doliwa, A., Grundland, A.M.: {Minimal surfaces in the soliton surface approach,
  arXiv ID: 1511.02173}  (2015)

\bibitem{Dubov1992}
Dubov, S.Y., Eleonskii, V.M., Kulagin, N.E.: {Equidistant spectra of anharmonic
  oscillators}.
\newblock Sov. Phys. JETP \textbf{75}(3), 446--451 (1992)

\bibitem{Dubov1994}
Dubov, S.Y., Eleonskii, V.M., Kulagin, N.E.: {Equidistant spectra of anharmonic
  oscillators}.
\newblock Chaos \textbf{4}(1), 47--53 (1994).
\newblock \doi{10.1063/1.166056}

\bibitem{enneper1868analytisch}
Enneper, A.: Analytisch-geometrische untersuchungen nachr.
\newblock K{\"o}nigl. Gesell. Wissensch. Georg--Augusts-Univ. G{\"o}ttingen
  \textbf{12}, 258--277 (1868)

\bibitem{Felder2012}
Felder, G., Hemery, A.D., Veselov, A.P.: {Zeros of Wronskians of Hermite
  polynomials and Young diagrams}.
\newblock Physica D: Nonlinear Phenomena \textbf{241}(23-24), 2131--2137
  (2012).
\newblock \doi{10.1016/j.physd.2012.08.008}

\bibitem{Filipuk2020}
Filipuk, G., Ishkhanyan, A., Derezi{\'{n}}ski, J.: {On the derivatives of the
  Heun functions}.
\newblock Journal of Contemporary Mathematical Analysis, Armenian Academy of
  Sciences (accepted for publication), ArXiv ID: 1907.12692 pp. 1--13 (2020)

\bibitem{Gomez-Ullate2014}
G{\'{o}}mez-Ullate, D., Grandati, Y., Milson, R.: {Rational extensions of the
  quantum harmonic oscillator and exceptional Hermite polynomials}.
\newblock Journal of Physics A: Mathematical and Theoretical \textbf{47}(1)
  (2014).
\newblock \doi{10.1088/1751-8113/47/1/015203}

\bibitem{Gomez-Ullate2018}
G{\'{o}}mez-Ullate, D., Grandati, Y., Milson, R.: {Durfee Rectangles and
  Pseudo-Wronskian Equivalences for Hermite Polynomials}.
\newblock Studies in Applied Mathematics \textbf{141}(4), 596--625 (2018).
\newblock \doi{10.1111/sapm.12225}

\bibitem{Gomez-Ullate2010}
G{\'{o}}mez-Ullate, D., Kamran, N., Milson, R.: {Exceptional orthogonal
  polynomials and the Darboux transformation}.
\newblock Journal of Physics A: Mathematical and Theoretical \textbf{43}(43)
  (2010).
\newblock \doi{10.1088/1751-8113/43/43/434016}

\bibitem{Gomez-Ullate2013}
G{\'{o}}mez-Ullate, D., Kamran, N., Milson, R.: {A Conjecture on Exceptional
  Orthogonal Polynomials}.
\newblock Foundations of Computational Mathematics \textbf{13}(4), 615--666
  (2013).
\newblock \doi{10.1007/s10208-012-9128-6}

\bibitem{Gomez-Ullate2016}
G{\'{o}}mez-Ullate, D., Kasman, A., Kuijlaars, A.B., Milson, R.: {Recurrence
  relations for exceptional Hermite polynomials}.
\newblock Journal of Approximation Theory \textbf{204}, 1--16 (2016).
\newblock \doi{10.1016/j.jat.2015.12.003}

\bibitem{Ho2012}
Ho, C.L., Odake, S., Sasaki, R.: {Zeros of the Exceptional Laguerre and Jacobi
  Polynomials}.
\newblock Symmetry, Integrability and Geometry: Methods and Applications
  (SIGMA) \textbf{7}(107), 1--27 (2012).
\newblock \doi{10.5402/2012/920475}

\bibitem{Hoffmann2018}
Hoffmann, S.E., Hussin, V., Marquette, I., Zhang, Y.Z.: {Non-classical
  behaviour of coherent states for systems constructed using exceptional
  orthogonal polynomials}.
\newblock Journal of Physics A: Mathematical and Theoretical \textbf{51}(8)
  (2018).
\newblock \doi{10.1088/1751-8121/aaa553}

\bibitem{Hoque2018}
Hoque, M.F., Marquette, I., Post, S., Zhang, Y.Z.: {Algebraic calculations for
  spectrum of superintegrable system from exceptional orthogonal polynomials}.
\newblock Annals of Physics \textbf{391}, 203--215 (2018).
\newblock \doi{10.1016/j.aop.2018.02.008}

\bibitem{Krein1957}
Krein, M.G.: {On a continuous analogue of a Christoffel formula from the theory
  of orthogonal polynomials}.
\newblock Dokl. Akad. Nauk SSSR \textbf{113}(5), 970--973 (1957)

\bibitem{Kuijlaars2015}
Kuijlaars, A.B., Milson, R.: {Zeros of exceptional Hermite polynomials}.
\newblock Journal of Approximation Theory \textbf{200}, 28--39 (2015).
\newblock \doi{10.1016/j.jat.2015.07.002}

\bibitem{Liaw2015}
Liaw, C., Littlejohn, L.L., Stewart, J., Wicks, Q.: {A spectral study of the
  second-order exceptional X1-Jacobi differential expression and a related
  non-classical Jacobi differential expression}.
\newblock Journal of Mathematical Analysis and Applications \textbf{422}(1),
  212--239 (2015).
\newblock \doi{10.1016/j.jmaa.2014.08.016}

\bibitem{Milson2019}
Milson, R.: {Toward the classification of Exceptional Orthogonal Polynomials: a
  progress report}.
\newblock In: Quantum Theory and Symmetries, CRM Proceedings and AMS Lecture
  Notes Ed. M. Paranjape (to appear) (2020)

\bibitem{Nikiforov1988}
Nikiforov, A.F., Uvarov, V.B.: {Special Functions of Mathematical Physics}.
\newblock Birkha{\"{u}}ser Verlag, Basel (1988).
\newblock \doi{10.1007/978-1-4757-1595-8}

\bibitem{Oblomkov1999}
Oblomkov, A.A.: {Monodromy-free Schr{\"{o}}dinger operators with quadratically
  increasing potentials}.
\newblock Theoretical and Mathematical Physics \textbf{121}(3), 374--386
  (1999).
\newblock \doi{10.1007/bf02557204}

\bibitem{Odake2013b}
Odake, S.: {Recurrence relations of the multi-indexed orthogonal polynomials}.
\newblock Journal of Mathematical Physics \textbf{54}(8) (2013).
\newblock \doi{10.1063/1.4819255}

\bibitem{Odake2010a}
Odake, S., Sasaki, R.: {Another set of infinitely many exceptional (Xl)
  Laguerre polynomials}.
\newblock Physics Letters, Section B: Nuclear, Elementary Particle and
  High-Energy Physics \textbf{684}(2-3), 173--176 (2010).
\newblock \doi{10.1016/j.physletb.2009.12.062}

\bibitem{Odake2013a}
Odake, S., Sasaki, R.: {Multi-indexed Wilson and Askey-Wilson polynomials}.
\newblock Journal of Physics A: Mathematical and Theoretical \textbf{46}(4)
  (2013).
\newblock \doi{10.1088/1751-8113/46/4/045204}

\bibitem{Ohyama2006}
Ohyama, Y., Okumura, S.: {A coalescent diagram of the Painlev{\'{e}} equations
  from the viewpoint of isomonodromic deformations}.
\newblock Journal of Physics A: Mathematical and General \textbf{39},
  12129--12151 (2006).
\newblock \doi{10.1088/0305-4470/39/39/S08}

\bibitem{Quesne2008}
Quesne, C.: {Exceptional orthogonal polynomials, exactly solvable potentials
  and supersymmetry}.
\newblock Journal of Physics A: Mathematical and Theoretical \textbf{41}(39)
  (2008).
\newblock \doi{10.1088/1751-8113/41/39/392001}

\bibitem{Quesne2012}
Quesne, C.: {Exceptional orthogonal polynomials and new exactly solvable
  potentials in quantum mechanics}.
\newblock Journal of Physics: Conference Series \textbf{380}(1) (2012).
\newblock \doi{10.1088/1742-6596/380/1/012016}

\bibitem{Samsonov1996}
Samsonov, B.F.: {New features in supersymmetry breakdown in quantum mechanics}.
\newblock Modern Physics Letters A \textbf{11}(19), 1563--1567 (1996).
\newblock \doi{10.1142/S0217732396001557}

\bibitem{Sasaki2010}
Sasaki, R., Tsujimoto, S., Zhedanov, A.: {Exceptional Laguerre and Jacobi
  polynomials and the corresponding potentials through Darboux-Crum
  transformations}.
\newblock Journal of Physics A: Mathematical and Theoretical \textbf{43}(31)
  (2010).
\newblock \doi{10.1088/1751-8113/43/31/315204}

\bibitem{Slavyanov2000}
Slavyanov, S.Y., Lay, W.: {Special Functions: A Unified Theory Based on
  Singularities}.
\newblock Oxford Science Publications, Oxford (2000)

\bibitem{Szego1939}
Szeg{\"{o}}, G.: {Orthogonal polynomials}.
\newblock American Mathematical Society, Providence (1939)

\bibitem{weierstrass1866fortsetzung}
Weierstrass, K.: Fortsetzung der untersuchung {\"u}ber die minimalflachen.
\newblock Mathematische Werke \textbf{3}, 219--248 (1866)

\end{thebibliography}

\end{document}